\documentclass[reprint,
superscriptaddress,
nofootinbib,
 amsmath,amssymb,
 aps,
pra,
floatfix,
]{revtex4-2}

\usepackage{xcolor} 
\usepackage{comment}
\usepackage{amsmath}  
\usepackage{amsfonts} 
\usepackage{graphicx} 
\usepackage[utf8]{inputenc}
\usepackage{pgfplots}
\DeclareUnicodeCharacter{2212}{?}
\usepgfplotslibrary{groupplots,dateplot}
\usetikzlibrary{patterns,shapes.arrows}
\pgfplotsset{compat=newest}
\usepackage{siunitx}
\usepackage{cancel}

\usepackage{hyperref}

\usepackage{tikz}

\usetikzlibrary{decorations.pathmorphing}
\tikzset{snakeit/.style={decorate, decoration={snake,amplitude=1.5pt, pre length=3mm,post length=1mm}}}

\usetikzlibrary{arrows.meta}
\usetikzlibrary{shapes.callouts}
\usetikzlibrary{shapes.multipart}
\tikzset{
  level/.style   = { ultra thick, blue },
  rlevel/.style   = { ultra thick, black },
  glevel/.style   = { ultra thick, gray },
  connect/.style = { dashed, black },
  notice/.style  = { draw, rectangle callout, callout relative pointer={#1} },
  label/.style   = { text width=2cm }
}

\renewcommand\[{\begin{equation}}
\renewcommand\]{\end{equation}}

\usepackage{pgf,tikz-3dplot}
\tikzset{%
glow/.style={%
preaction={#1, draw, line join=round, line width=0.5pt, opacity=0.04,
preaction={#1, draw, line join=round, line width=1.0pt, opacity=0.04,
preaction={#1, draw, line join=round, line width=1.5pt, opacity=0.04,
preaction={#1, draw, line join=round, line width=2.0pt, opacity=0.04,
preaction={#1, draw, line join=round, line width=2.5pt, opacity=0.04,
preaction={#1, draw, line join=round, line width=3.0pt, opacity=0.04,
preaction={#1, draw, line join=round, line width=3.5pt, opacity=0.04,
preaction={#1, draw, line join=round, line width=4.0pt, opacity=0.04,
preaction={#1, draw, line join=round, line width=4.5pt, opacity=0.04,
preaction={#1, draw, line join=round, line width=5.0pt, opacity=0.04,
preaction={#1, draw, line join=round, line width=5.5pt, opacity=0.04,
preaction={#1, draw, line join=round, line width=6.0pt, opacity=0.04,
}}}}}}}}}}}}}}
\usetikzlibrary{decorations.pathmorphing}

\begin{document}

\global\long\def\normsq#1{\left|#1\right|^{2}}%
\global\long\def\ket#1{\left|#1\right\rangle }%
\global\long\def\bra#1{\left\langle #1\right|}%
\global\long\def\abs#1{\left|#1\right|}%
\global\long\def\normsq#1{\left|#1\right|^{2}}%
\global\long\def\mean#1{\left\langle #1\right\rangle }%
\global\long\def\braket#1#2{\left\langle #1|#2\right\rangle }%
\global\long\def\ketbra#1#2{\left|{#1}\rangle\!\langle{#2}\right|}%
\global\long\def\ketbrad#1{\left|{#1}\rangle\!\langle{#1}\right|}%
\global\long\def\dt{\frac{d}{dt}}%
\global\long\def\pprime{\prime\prime}%
\global\long\def\bbZ{\mathbb{Z}}%
\global\long\def\bbR{\mathbb{R}}%
\global\long\def\tt#1{\mathtt{#1}}%


\title{Bursts of polarised single photons from atom-cavity sources}

\author{Jan Ole Ernst}
\affiliation{Clarendon Laboratory, University of Oxford, Parks Road, Oxford OX1 3PU, United Kingdom}

\author{Juan-Rafael Álvarez}
 \altaffiliation[Now at ]{Université Paris-Saclay, CNRS, Centre de Nanosciences et de Nanotechnologies, 91120, Palaiseau, France}

\author{Thomas D. Barrett}
\altaffiliation[Now at ]{InstaDeep, London, UK.}

\author{Axel Kuhn}
\email{axel.kuhn@physics.ox.ac.uk}
\affiliation{Clarendon Laboratory, University of Oxford, Parks Road, Oxford OX1 3PU, United Kingdom}




\begin{abstract}

Photonic qubits play an instrumental role in the development of advanced quantum technologies, including quantum networking, boson sampling and measurement based quantum computing. A promising framework for the deterministic production of indistinguishable single photons is an atomic emitter coupled to a single mode of a high finesse optical cavity. Polarisation control is an important cornerstone, particularly when the polarisation defines the state of a quantum bit. Here, we propose a scheme for producing bursts of polarised single photons by coupling a generalised atomic emitter to an optical cavity, exploiting a particular choice of quantisation axis. In connection with two re-preparation methods, simulations predict 10-photon bursts coincidence count rates on the order of 1 kHz with single $^{87}\mathrm{Rb}$ atoms trapped in a state of the art optical cavity. This paves the way for novel $n$-photon experiments with atom-cavity sources.

\end{abstract}

\maketitle 

\section{Introduction}

Photonic qubits are key to many quantum technologies such as linear optical \cite{knillSchemeEfficientQuantum2001} and measurement-based \cite{mbqc} quantum computing, as well as boson sampling \cite{aaronsonComputationalComplexityLinear2011}. Moreover, photonic qubits are important for the development of quantum networks \cite{ritterElementaryQuantumNetwork2012}, where the state of a stationary qubit is coherently mapped to a photonic qubit for communication and entanglement distribution. Single photons are also ubiquitous in quantum metrology \cite{single_photon_sources_review}. What makes photons appealing is the ease with which they can be manipulated and transmitted with standard optical components, as well as their robustness to decoherence.

The implementation of optical quantum technologies requires highly efficient single photon sources. Spontaneous parametric down conversion sources have been used widely and successfully in this regard \cite{Zhang2021}, yet their probabilistic nature make generating larger photon sequences and larger entangled states \cite{Yao2012}, required for scaleable quantum computing architectures, challenging. To overcome the scalability challenges associated with an inherently probabilistic architecture, deterministic schemes for single photon generation have been proposed and shown promise, such as quantum dots \cite{Istrati2020}, ions \cite{Keller2004, northup_indistinguishable_photons} or neutral atoms \cite{kuhn2002}.   

 Many deterministic single photon production schemes exploit the coupling of a quantum emitter to an optical resonator, allowing for the adiabatic transfer of atomic population between two stable ground states \cite{predojevicEngineeringAtomPhotonInteraction2015} in a process known as Vacuum-enhanced Stimulated Raman Adiabatic Passage (V-STIRAP) \cite{kuhn2002}. Photons produced coherently with Raman laser pulses and optical cavities can have very long coherence times, on the order of a few hundreds of nanoseconds \cite{nisbet-jonesPhotonicQubitsQutrits2013}. The advantages of this technique are the precise control over all photonic degrees of freedom, including polarisation, frequency and wave-packet shape \cite{vasilevSinglePhotonsMadetomeasure2010}. This is promising as it allows for the integration of such sources with various other physical platforms, such as multimode interferometers 
 \cite{barrettMultimodeInterferometryEntangling2019} or quantum networks for the remote entanglement of distant atomic emitters. As the photons produced are highly indistinguishable, they can also be used to explore foundational questions in quantum optics about the nature of the photon as a fundamental excitation \cite{antidote}. Cavity-QED platforms have been used succesfully for quantum information processing, such as the implementation of an optical C-NOT gate \cite{holleczekQuantumLogicCavity2016}, the demonstration of an efficient quantum repeater \cite{rempe_repeater} or the efficient generation of large entangled cluster states \cite{rempe_cluster}.

Nevertheless, atom-cavity architectures are subject to various limitations. Often, the rate at which single photons are produced coherently from a deterministic atomic source is far lower than that of many non-deterministic sources, making it challenging to implement quantum information processing protocols that require large photon numbers or reasonable clock rates. Moreover, polarisation control is hard to achieve due to a manyfold of degenerate magnetic substates in most atoms. To overcome the latter, historically the production of polarisation controlled single photons relied on applying strong external magnetic fields to individually address a single Zeeman-state \cite{Wilk_2007a,Wilk_2007b}. However, the physical cavity parameters demand a rather strong magnetic field to lift the degeneracy of states, which led to the apparition of nonlinear Zeeman effects \cite{barrettNonlinearZeemanEffects2018}, strongly curtailing the achievable efficiencies of such schemes. 

In this paper, we propose a cavity-based scheme to generate bursts of polarisation controlled single photons which does not require a strong external magnetic field and enables the fast and efficient coherent re-preparation of the atom back into the desired initial state. The implementation in an idealised as well as a real atomic system, where a single neutral atom is loaded and trapped in a high-finesse cavity, is explored extensively in simulation. This paper is structured as follows: in Sec. \ref{sec:ideal} the implementation of this scheme in an idealised atomic system is outlined, which encompasses the production of polarised photons, as well as the coherent re-preparation of the atom into the desired initial state. In Sec \ref{sec:rb87}, we investigate the  feasibility of implementing the using $^{87}\mathrm{Rb}$, which contains two particular transitions that describe the required energy level structure. We discuss various trade-offs and limitations with respect to a variety of cavity parameters. Additionally, we analyse the efficiency limitations of the scheme and present an optimal sequence which uses a combination of incoherent and coherent state preparation methods. Finally, Sec. \ref{sec:outlook} contextualises our results, and illustrates potential future directions.

\section{Implementation in idealised three level system}
\label{sec:ideal}

\begin{figure}
\begin{tikzpicture}[decoration = {snake,pre=lineto,pre length=0.1cm,post=lineto,post length=0.1cm},scale=1.1, every node/.style={transform shape}]

\draw[black, ] (1.5,-0.5) -- (2.5,-0.5) node[right]{$\ket{g_{2}}$} node[right,xshift=2.5cm]{$F+1$};

\draw[gray, dashed] (1.5,-1.2) -- (2.5,-1.2);

\draw[black, ] (0,-1.2) node[left]{$\ket{g_{1}^{-}}$} -- (1,-1.2);
\draw[black, line width=0.5] (3,-1.2) --  (4,-1.2) node[right]{$\ket{g_{1}^{+}}$} node[right, xshift=1cm]{$F$};

\draw[black, ] (1.5,2) -- node[above]{$\ket{x}$} (2.5,2) node[right,xshift=2.5cm]{$F^{\prime} = F$};

\draw[black, dashed] (0,2) -- (1,2);
\draw[black, dashed] (3,2) -- (4,2);

\draw[-stealth,black,thick] (2.1,-0.5) -- node[right, yshift=-0.2cm,xshift=-0.08cm,scale=0.85]{$\Omega_{\mathrm{VST}}$} node[red,right,xshift=-0.08cm,yshift=-0.5cm,scale=0.85]{$\pi$} (2.1,2);

\draw[red,opacity=0.5] (1.9,-1.2) --  node[red,scale=0.85]{$\times$} (1.9,2);

\draw[stealth- ,dashed] (0.4,-1.1) -- node[left,xshift=-0.05cm,scale=0.85]{$g$} node[red,left,xshift=-0.15cm,yshift=-0.35cm,scale=0.85]{$\sigma^{+}$} (1.8,2);

\draw[stealth-, dashed ] (3.6,-1.1) -- node[right,xshift=0.05cm,scale=0.85]{$g$}  node[red,right,xshift=0.3cm,yshift=-0.35cm,scale=0.85]{$\sigma^{-}$}(2.2,2);

\draw[blue,glow=blue] (3.6,-1.2)  arc
	[
		start angle=0,
		end angle=-180,
		x radius=1.6cm,
		y radius =0.4cm
	] node[below,xshift=0.5cm,yshift=-0.4cm]{$\ket{\Phi_{-}}$}  ;

\end{tikzpicture}

    \caption{Idealised system with three atomic energy levels: two stable ground levels with total angular momenta $F,F+1$ and one excited level with total angular momentum $F'=F$. The individual states $\ket{g_1^{-}}, \ket{g_1^{0}}, \ket{g_1^{+}}$ are degenerate. The excited state $\ket{x}$ has dipole allowed transitions to $\ket{g_1^{\pm}}$ and $\ket{g_2}, $ and the $\pi$ transition to $\ket{g_1^0}$ is dipole forbidden. The same configuration can of course arise for atoms with no hyperfine structure and $F=J$. This configuration of levels is required for the polarised photon production scheme.}
    \label{fig:ideal_3lvl}
\end{figure}
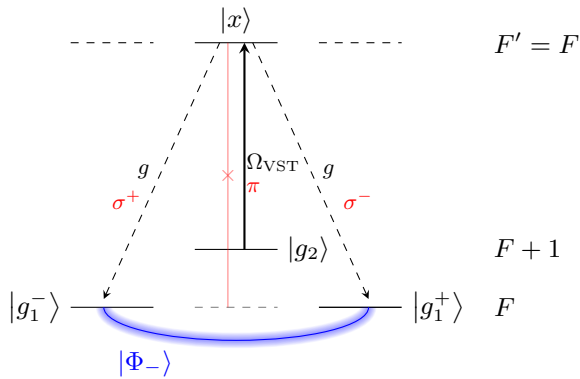

\subsection{Atomic Structure}
We first present our proposal for polarisation controlled single photon production in an idealised three level system. 

The atomic structure required for the production of polarised photons comprises  two ground levels with total angular momenta $F$ and $F+1$, as well as one excited level with total angular momentum $F'=F$. Three of the individual states of the two ground levels are required for photon production, which we call $\{ \ket{g_2}, \ket{g_1^+}, \ket{g_1^-} \}$. The latter two levels can be considered to be degenerate in the absence of a magnetic field. The excited state, $\ket{x}$, has dipole-allowed transitions to all three ground states (cf. Fig \ref{fig:ideal_3lvl}), yet the transition to $\ket{g_1^0}$ is dipole forbidden. A transition from $\ket{x}$ to either of the degenerate ground states $\ket{g_1^{\pm}}$ corresponds to the emission of photons of orthogonal circular polarisations $\sigma^{\mp}$ with respect to the chosen quantisation axis, while the transition $\ket{g_2} \to \ket{x}$ corresponds to the absorption of a photon of $\pi$-polarisation. This configuration arises naturally in alkali atoms and ions, both with and without hyperfine structure (for example $\mathrm{Rb}$, $\mathrm{Cs}$, $\mathrm{Ca}^{+}$, $\mathrm{Sr}^{+}$). Nevertheless, careful considerations with respect to the precise choice of atom must me made (cf. Sec. \ref{sec:rb87}). \\

 The cornerstones of our scheme are a proper choice of quantisation axis and polarisation directions for the atomic driving, as well as a fast coherent re-preparation of the atom back to its original state. This enables the production of a burst of polarised single photons emitted with repetition rates on the MHz scale. Optical pumping can be used to initialise the atom at the outset and to re-initialise it after failed photon production. In what follows, we describe these cornerstones, as well as the implementation of the overall sequence.

\subsection{Quantisation axis and polarised photon production}
\label{sec:quantax}

\begin{figure}
\begin{centering}

\includegraphics[width=\columnwidth]{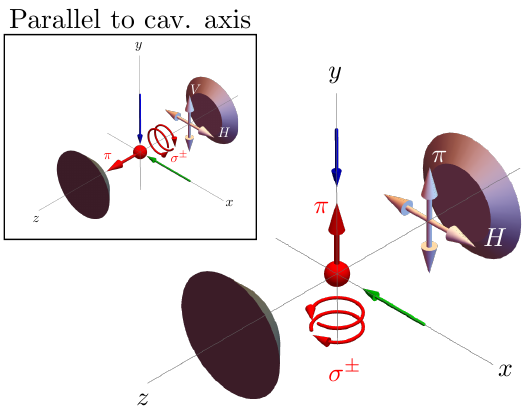}

\par\end{centering}
\caption{Atomic Quantisation axis perpendicular to the cavity axis. In red the $\pi$, $\sigma^{+}$ and $\sigma^{-}$ polarisation basis with respect to this particular quantisation axis is shown. In  green, the light addressing the atom for V-STIRAP is shown. In white, we see the two linear, orthogonal polarisation modes that span the cavity basis. The inset shows the traditional choice of quantisation axis, parallel to the cavity axis, where the cavity supports either a circularly polarised basis $\{ \sigma^+, \sigma^- \}$ or a linear basis $\{ H, V\}$. }
\label{fig:cav_quant}
\end{figure}

An important aspect of our proposal is a choice of quantisation axis which is perpendicular to the cavity axis as shown in Fig. \ref{fig:cav_quant}.  In combination with the atomic selection rules, this uniquely defines the polarisation of the photon produced by the atom-cavity interaction.

The cavity supports two degenerate polarisation modes, one is $\pi$-polarised and the other a linear superposition of $\sigma^{+}$ and  $\sigma^{-}$. Without loss of generality, for a quantisation axis along the $y$ axis, the polarisation modes supported by the cavity correspond to $\{\pi, H\}$ (in the lab frame, these correspond to polarisations parallel to the $\{x, y\}$ axes, cf. Fig. \ref{fig:cav_quant}). We aim to emit only into the linear polarisation modes supported by the cavity by enabling the atom to emit photons in a superposition state with respect to the quantisation axis, following ideas first introduced in \cite{wilkSingleAtomSinglePhotonQuantum2007} for $H$ polarised photons from atom-photon entanglement.

A single atom is initially prepared in the state $\ket{g_2}$ and addressed by a laser with $\pi-$polarisation and total pulse duration $T$ (polarised in the $y$ direction in the lab frame, propagating along $x$ cf. Fig. \ref{fig:cav_quant}), resonant with the transition from $\ket{g_2}$ to $\ket{x}$. The cavity is tuned to be resonant with the $\ket{g_1^{\pm}} \to \ket{x}$ transitions, supporting the dipole allowed $\sigma^{\mp}$ transitions while there is no dipole allowed $\pi$ transition. The cavity therefore only supports the $H = (\sigma^{+} + \sigma^{-})/ \sqrt{2}$ polarisation mode. This paves the way for efficiently producing linearly polarised photons without a strong external magnetic field to address individual hyperfine states.

\subsection{Adiabatic transfer of population} \label{sec:vst}

To produce photons, we consider a process known as Vacuum-induced Stimulated Raman Adiabataic Passage (V-STIRAP) \cite{VSTIRAP2000} whereupon the atomic population is adiabatically transferred from the initial state $\ket{g_2}$ to a superposition of states $\ket{g_1^{\pm}}$, leading to the following entangled atom-photon state, $\ket{\Psi}$:
\begin{equation}
\begin{aligned}
    \ket{\Psi}&=\frac{1}{\sqrt{2}}\left( \ket{g_1^+, \sigma^-} - \ket{g_1^-, \sigma^+} \right) \\ &=\frac{1}{\sqrt{2}}\left(\ket{\Phi_{\phi=\pi}}\otimes\ket{H}-i\ket{\Phi_{\phi=0}}\otimes\ket{V} \right),
\label{eq:gs-sup}
\end{aligned}
\end{equation}
where $\ket{\Phi_{\phi}}=\frac{1}{\sqrt{2}}\left(\ket{g_1^+}+e^{i\phi}\ket{g_1^-} \right)$ \footnote{In general, the phase $\phi$ depends on the choice of quantisation axis and it should be noted that it is an important aspect of the coherent re-preparation scheme described in Sec. \ref{sec:ideal_cr}.}.
However, the cavity only supports one linear polarisation mode $H=(\sigma^{+} + \sigma^{-})/ \sqrt{2}$, where $H$ is perpendicular to the cavity axis. $V$ would correspond to an oscillation along the cavity axis and photons propagating perpendicular to the cavity axis. Therefore, the cavity acts like a polarisation filter, only supporting $H$-polarised light. As a consequence, the atom-cavity state collapses to the product state: \begin{equation}
    \ket{\Psi_{H}}=\ket{\Phi_{\phi=\pi}}\otimes\ket{H},
\label{eq:f1sup0}
\end{equation}
which, upon photon leakage from the cavity, evolves to \begin{equation}
\ket{\Psi_{f}}=\ket{\Phi_{\phi=\pi}}\otimes\ket{0}.\label{eq:f1sup}
\end{equation}

\subsection{Efficiency} \label{sec:efficiency}

The efficiency of the V-STIRAP process is determined by estimating the cavity field decay rate with respect to the desired final state, given by an integral over the photon production time interval $[t_i, t_f]$:  \begin{equation}
\label{eq:eff_1}
\eta = 2\kappa \int_{t_{i}}^{t_f} \braket{\Phi_{\pi}}{\rho(t)} \braket{\rho(t)}{\Phi_{\pi}} dt,
\end{equation}
In the idealised state configuration, there is an equivalent analytic formulation which does not require an integral (cf. App. \ref{sec:app_eff} for details). 

\begin{figure}[ht]
    \centering
    \includegraphics[width=\columnwidth]{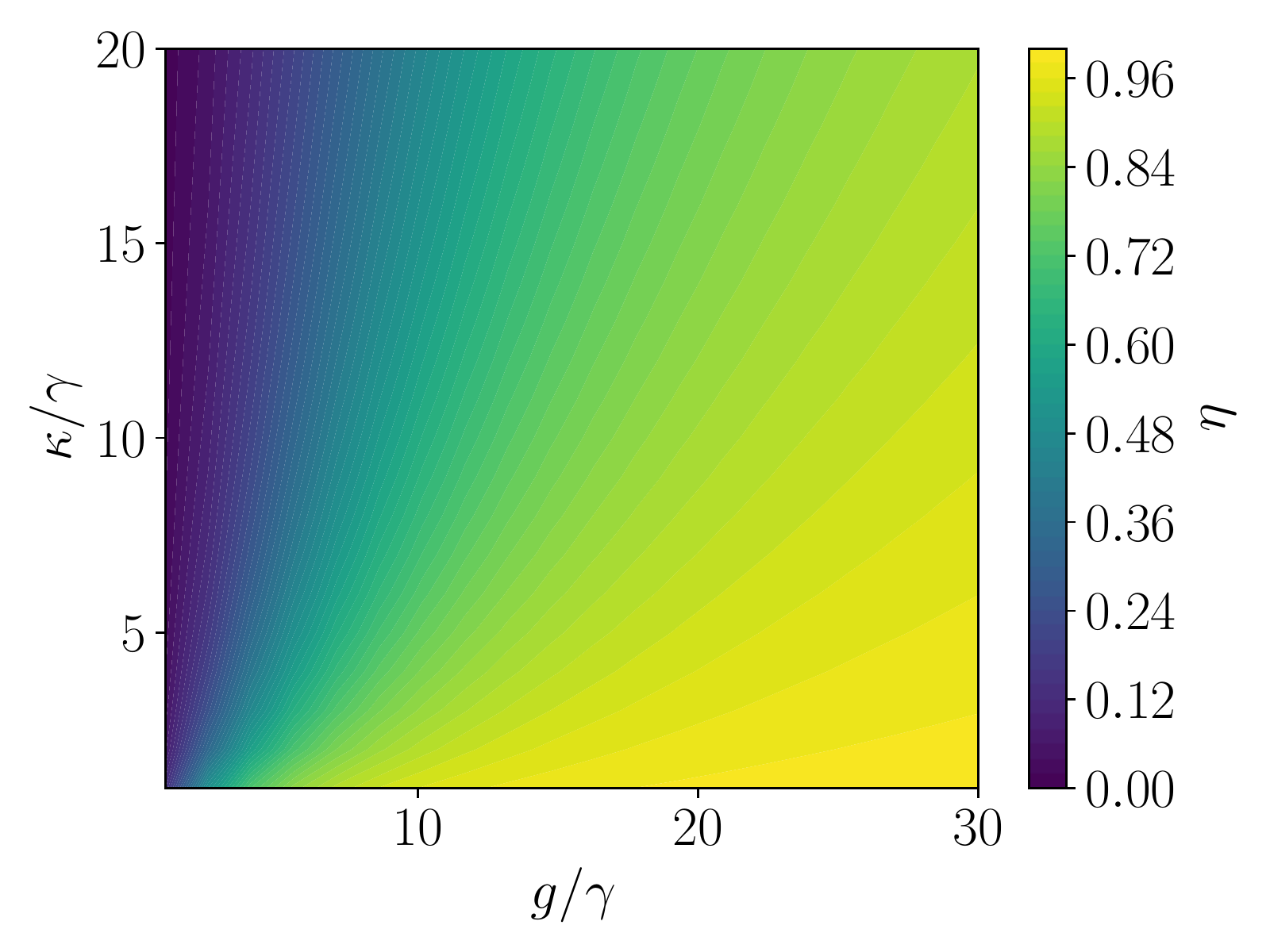}
    \caption{V-STIRAP photon production efficiency in an idealised system plotted for a fixed pulse of $T=10 / \gamma$ as a function of $\kappa / \gamma $ where $\kappa$ is the cavity field decay rate, as well as for different choices of $ g / \gamma$ where $g$ is the atom cavity coupling strength and $\gamma$ the atomic depolarisation rate. One can see a clear increase in efficiency for more strongly coupled systems. Each data-point corresponds to a local optimisation problem for finding the optimal peak Rabi frequency $\Omega_{\mathrm{max}}$ which maximises the photon production probability (cf. App. \ref{sec:app_ideal_vst} for further details). }
    \label{fig:vst_ideal_length}
\end{figure}

We calculate various numerical photon production efficiencies in an idealised state configuration in Fig. \ref{fig:vst_ideal_length}, by assuming an atom is successfully initialised in $\ket{g_2}$ and addressed with a sinusoidal laser pulse of $\Omega_{\mathrm{VST}}= \Omega_{\mathrm{max}}\sin^2(\pi t / T )$, where the total pulse duration $T=10 / \gamma$. The variation as a 
     function of $\kappa/\gamma$, as well as $ g/ \gamma$, where $\kappa$ is the cavity field decay rate, $g$ the atom cavity coupling strength  \footnote{$g = g_{\mathrm{max}}* d_{\mathrm{i,f}}$, where $g_\mathrm{max}$ is the maximal atom cavity coupling $\propto 1 / \sqrt{V} $ where $V$ is the cavity mode volume and $d_{\mathrm{i,f}}$ is the Clebsch-Gordan coefficient for the $\sigma^{\mp}$ transitions from $\ket{g_1^{\pm}}$ to $\ket{x}$. } and $\gamma$ the atomic depolarisation rate are shown in Fig. \ref{fig:vst_ideal_length}. Each data-point corresponds to a local optimisation problem for finding the optimal peak Rabi frequency $\Omega_{\mathrm{max}}$ which maximises the photon emission efficiency (cf. App. \ref{sec:app_ideal_vst} for further details, where all the exact parameters are detailed).

Physically, the constraints which limit the efficiency of photon production are two-fold: firstly, the need for adiabaticity in the population transfer process and second, the ability to extract the photons produced in the cavity. The strength of the atom cavity coupling $g$ in relation to the atomic depolarisation rate $\gamma$, as well as the time-scale of the driving pulse $\Omega(t)$ with respect to the Rabi frequency $\Omega$ determine the extent to which the process can remain adiabatic. Moreover, the rate at which the photonic field decays, $\kappa$, is instrumental for extracting the photon efficiently from the cavity. As evident in Fig. \ref{fig:vst_ideal_length} the efficiency initially increases in proportion to $\kappa$ and approaches unity for higher $g$ once $ T \kappa >1$, because the extraction from the cavity no longer limits the efficiency of the photon production process. For shorter pulse durations, the process is also limited by a lack of adiabaticity and an increased likelihood of decoherence. We observe an efficiency trade-off exhibited by the photon production process: the desire for higher repetition rates competes with the efficiency, which will be discussed more extensively for a real atom in Sec. \ref{sec:discussion}.    \\

A fundamental upper bound in the photon production efficiency obtained for V-STIRAP arises from the atom-cavity parameters. Controlling the time evolution of the Rabi frequency $\Omega$ enables the precise shaping of the time-dependent photon amplitude \cite{vasilevSinglePhotonsMadetomeasure2010}, as well as the efficiency with which it is produced. Even with lossless mirrors, the production of a temporally shaped photon is limited in efficiency by a theoretical maximum of  $\eta = 2C/ (2C+1)$ \cite{vasilevSinglePhotonsMadetomeasure2010,predojevicEngineeringAtomPhotonInteraction2015}, where $C=g^2/(2\kappa \gamma)$ is the cooperativity of the cavity. 

\subsection{STIRAP Re-preparation}
\label{sec:ideal_cr}
 Coherent STIRAP re-preparation \cite{Bergmann2019} enables fast and efficient re-preparation of the atom back into the desired initial state $\ket{g_2}$ with high re-preparation efficiencies. Unlike incoherent methods, which rely on eventual spontaneous emission into the desired state and are bounded by the atomic decay rate, coherent repumping over a time-scale $T$ must satisfy the constraint of global adiabaticity \cite{Bergmann2019}, $\Omega_{\mathrm{rms}} T \gg 1$, where $\Omega_{\mathrm{rms}}=\sqrt{\hat{\Omega}_S^2+\hat{\Omega}_P^2} $, such that $\hat{\Omega}$ describes the time average amplitude over the pulse duration for the Stokes pulse $\Omega_{S}$ and the Pump pulse $\Omega_{P}$. The desire for short pulses with high repetition rates can be achieved with intense pulses with correspondingly high Rabi frequencies. Moreover, fully coherent atomic population transfer is reversible, 
 essential for the implementation of quantum networks \cite{ciracQuantumStateTransfer1997,dilleySinglephotonAbsorptionCoupled2012} and for the preparation of highly entangled optical states with a single atomic memory \cite{rempe_cluster}.
 
 After a photon has been successfully emitted, the atom, which is left in the state $\ket{\Phi_{\pi}}$, can be prepared back into the desired initial state $\ket{g_2}$ using a two pulse STIRAP \cite{stirap_review} procedure which involves using two laser pulses (Stokes and Pump) to transfer the atomic population back to its initial state. Coherent methods like STIRAP are generally extremely good for achieving high population transfer efficiencies \cite{vasilevOptimumPulseShapes2009,raman-exact-laforgue} and are also fast compared to incoherent optical pumping.
 
To perform coherent repumping, the pump pulse must be resonant with the transitions from the groundstates $\ket{g_{1}^{+}}$ and $\ket{g_{1}^{-}}$, to $\ket{x}$ and a $\pi$ polarised Stokes pulse must be resonant with respect to the $\ket{g_2} \rightarrow \ket{x}$ transition \footnote{Compared to the V-STIRAP process, the initial and final states swap roles and a pump laser and stimulating cavity mode are replaced by a stokes and pump laser respectively.}. The pump pulse must be polarised in a linear combination of $\sigma^+$ and $\sigma^-$ polarisation with respect to the quantisation axis, maintaining a particular phase relation between the circularly polarised components(cf. Eq. \ref{eq:f1sup}), such that it is linearly polarised perpendicularly both to the cavity and the quantisation axis \footnote{This requires the laser to be pointing along the quantisation axis, as explained in Fig. \ref{fig:exp-sequence}}. As a result, the polarisation of the laser light and thus the phase between the two circularly polarised components of the light must match the phase of photon polarisation derived from Eq. \eqref{eq:f1sup}. Both laser pulses must have equal effective Rabi frequencies \footnote{It is the effective peak Rabi frequencies which are equal. The absolute laser intensities may differ if the Clebsch-Gordan coefficients of the individual transitions vary.}.

\begin{figure}[ht]
    \centering
    \includegraphics[width=\columnwidth]{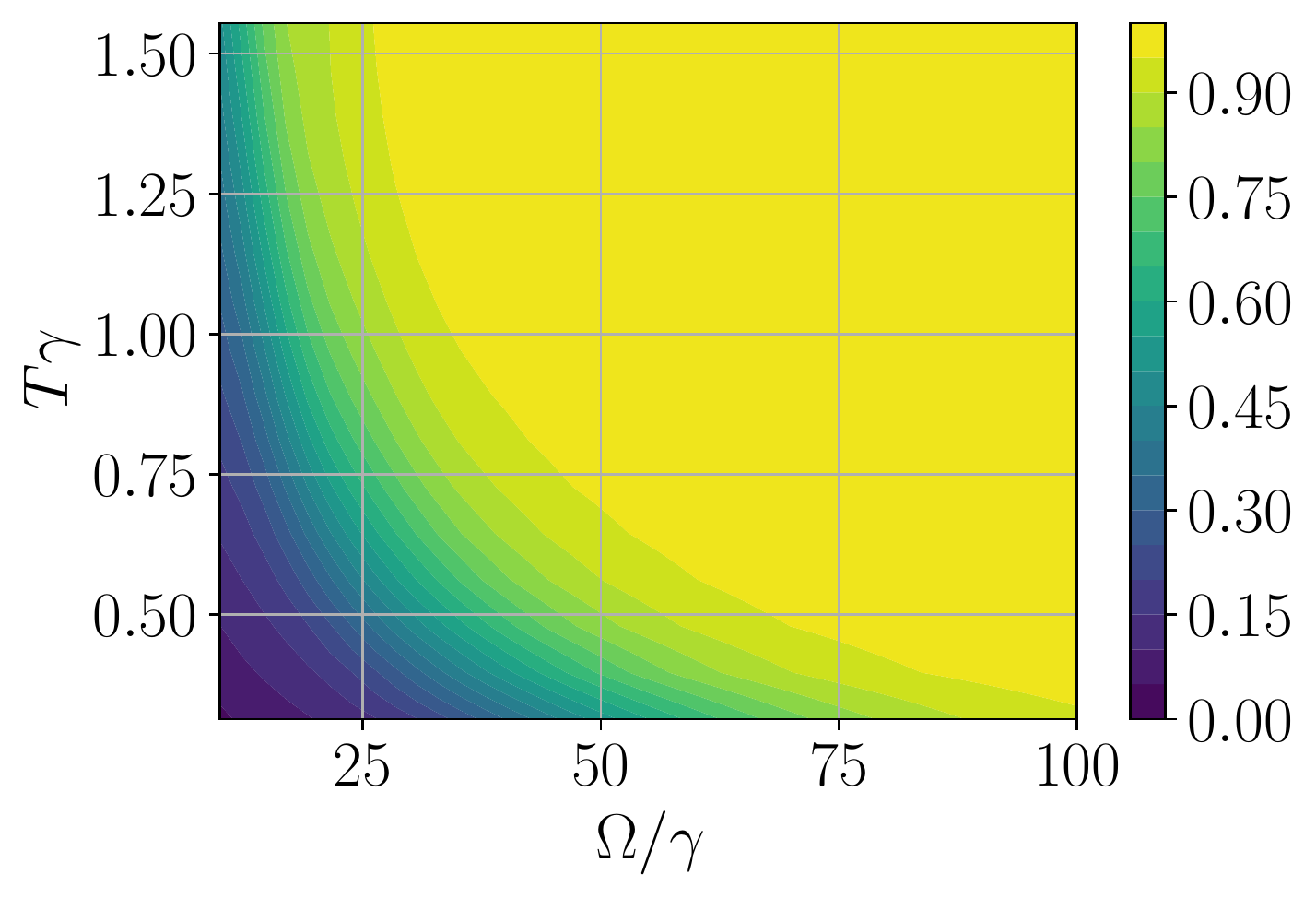}
    \caption{Repumping efficiency in idealised system plotted for length of the repumping pulses $T$ and peak Rabi frequency $\Omega$, which are both normalised with respect to the atomic depolarisation rate $\gamma$.}
    \label{fig:rep_eff_ideal}
\end{figure}

By carefully shaping the time dependent amplitudes of two pulses which address the transitions $\ket{g_1^{\pm}} \to \ket{x}$ and $\ket{x} \to \ket{g_2}$, it is possible to  maximise re-preparation efficiency between the states $\ket{g_1^{\pm}}$ and $\ket{g_2}$. Various pulse-shaping techniques have been developed extensively, such as optimal control methods \cite{stirap_oc_2002, nmr_oc, raman-exact-laforgue}, reinforcement learning \cite{GIANNELLI2022128054} or other numerical methods to minimise the non-adiabatic transitions \cite{vitanov_2009}. It should be noted that for experimentally feasible pulse shapes, one requires $\Omega_{S}(t=0)=\Omega_{P}(t=0)=\Omega_{S}(t=T)=\Omega_{P}(t=T)=0$, where $T$ is the total pulse sequence duration. \footnote{It is worth noting that if the transition used for V-STIRAP photon production corresponds to that of the Stokes laser (i.e., $\Omega_S$), the laser can be left on after the photon has been succesfully emmited from the cavity and turned off once the Pump pulse begins to increase in amplitude.} With this in mind, the highest population transfer efficiency was attained by adopting a technique that involved employing two interleaved pulses of constant rms Rabi frequency $\Omega_{\mathrm{rms}}$, which were applied with a hyper-Gaussian mask \cite{vitanov_2009}. The shape of the mask was numerically optimised for maximal population transfer efficiency (for more details cf. App. \ref{sec:app_ideal_rep}). The simulated STIRAP re-preparation efficiency approach unity in the idealised system configuration ($\eta_{\mathrm{max}} \approx 0.997$ in Fig. \ref{fig:rep_eff_ideal}), with losses parameterised by $\gamma$, as shown in Fig. \ref{fig:rep_eff_ideal}.

There is a trade-off however, between efficiency, repetition rate and driving laser intensity. The variation of the repumping efficiency with re-preparation time and peak Rabi frequencies is shown in Fig. \ref{fig:rep_eff_ideal}. In general, faster re-preparation requires larger peak Rabi frequencies and in the idealised system configuration one can in principle arbitrarily increase the peak Rabi frequency to increase the efficiency, this is in stark contrast to many real atomic systems where these are limited.

\section{Real atomic system: Rubidium 87}
\label{sec:rb87}
Many alkali atoms and earth alkaline ions exhibit the desired atomic structure. A particular choice for an atomic emitter is $^{87}\mathrm{Rb}$, as its properties and atomic structure are well understood \cite{steck87} and it exhibits a hyperfine structure suitable for the proposal. Following the exposition in Sec. \ref{sec:vst}, one picks $\ket{g_1^{\pm}} = \ket{F=1, m_F=\pm 1}$, $\ket{g_2} = \ket{F=2, m_F=0}$, and $\ket{x} = \ket{F'=1, m_F=0}$, . This configuration of states in Rubidium exists for two different transitions, one has the choice of either the $D_2$ transitions defined with respect to $5^2S_{\frac{1}{2}} \Longleftrightarrow 5^2P_{\frac{3}{2}}$ or the $D_1$ transitions, defined with respect to $5^2S_{\frac{1}{2}} \Longleftrightarrow 5^2P_{\frac{1}{2}}$. In what ensues, all components of the photon production sequence will be analysed and the ideal parameters are presented. An important trade-off with respect to photon production efficiency and the explicit cavity and laser parameters are discussed in Sec \ref{sec:ph_prod} and the results will be discussed and presented in Sec. \ref{sec:discussion}.
\begin{center}
\begin{figure*}
\begin{centering}

\includegraphics[width=2\columnwidth]{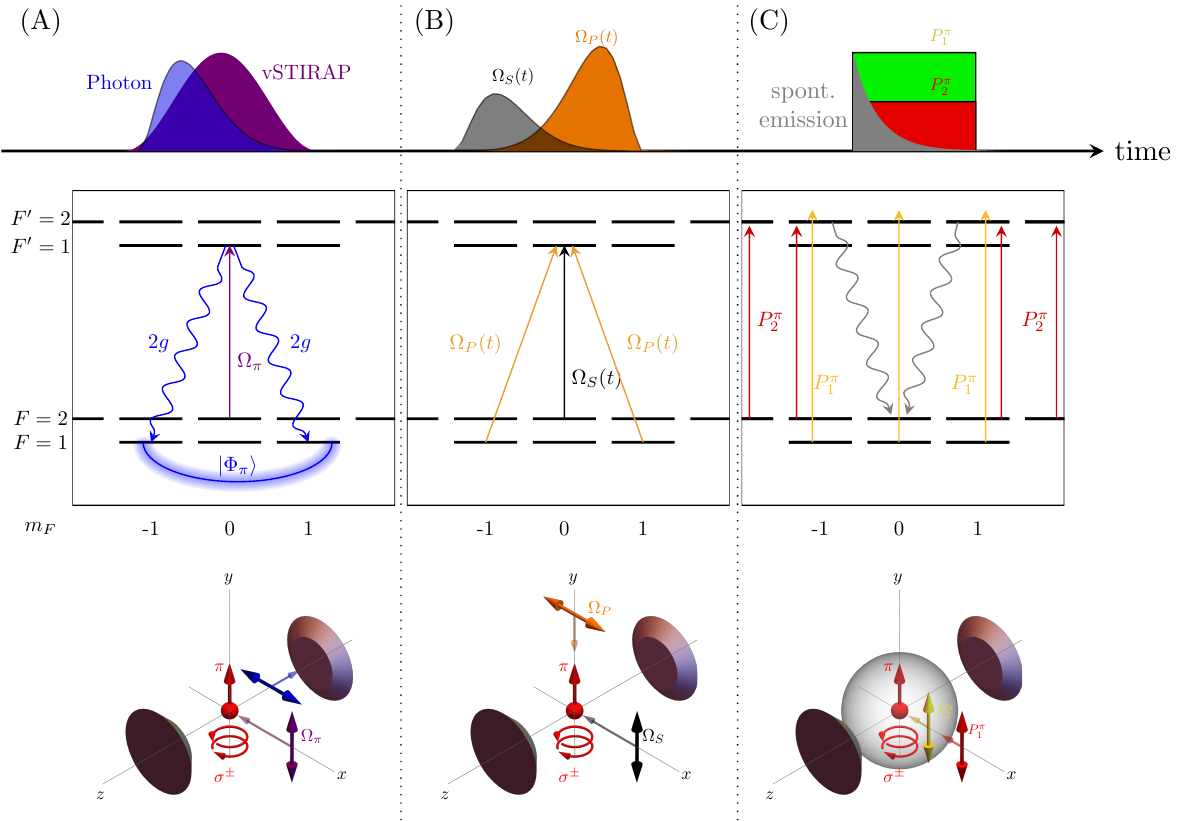}

\par\end{centering}
\caption{Sequence for photon generation with $^{87}\mathrm{Rb}$. (A) Starting with an atom in $\ket{F=2, m_F=0}$, a $\pi$ polarised driving pulse $\Omega_{\pi}$, resonant with the transition to $\ket{F'=1, m_{F'}=0}$, addresses the atom. The cavity is resonant with the transitions $F=1 \leftrightarrow F'=1$. Importantly, the $\pi$-transition from $\ket{F'=1, m_{F'}=0} \rightarrow \ket{F=1, m_{F}=0} $ is dipole forbidden such that the atomic population is transferred into the superposition state $\ket{\Phi_{\phi=\pi}}=\frac{1}{\sqrt{2}}(\ket{F=1, m_F=1}-\ket{F=1, m_{F}=-1})$. A linearly polarised photon is then emitted. (B) This is followed by fast, coherent STIRAP re-preparation. A driving laser of peak Rabi frequency $\Omega_{P}$ is polarised linearly and parallel to the emitted photon and a stimulating laser $\Omega_{S}$ that is $\pi$ polarised. Together, they adiabatically re-prepare the atom back into the desired initial state $\ket{F=2, m_F=0}$. These two processes can be repeated up to 10 times to produce the desired number of photons sequentially. (C) Thereafter, the atom is re-prepared incoherently into the desired initial state with optical pumping. One laser is blue detuned from the $F=1 \rightarrow F'=2$ transition and the other is red detuned from the $F=2 \rightarrow F'=2$ transition. All transitions from the excited $F'=2$ line are dipole allowed except the $\ket{F'=2, m_{F'}=0} \rightarrow \ket{F=2, m_{F}=0}$ transition. The lasers are two overlapping top-hat pulses, thus eventually the population is continually re-excited out of the undesired Zeeman ground-states and will spontaneously decay (isotropically, shown as a grey sphere) to the desired $\ket{F=2, m_F}=0$ state (cf. grey lines). This is also used to initialise the atom into the desired Zeeman state in the first place.}
\label{fig:exp-sequence}
  \end{figure*}
\par\end{center}

\subsection{Photon Production}
\label{sec:ph_prod}
\begin{figure}[ht]
    \centering
    \includegraphics[width=\columnwidth]{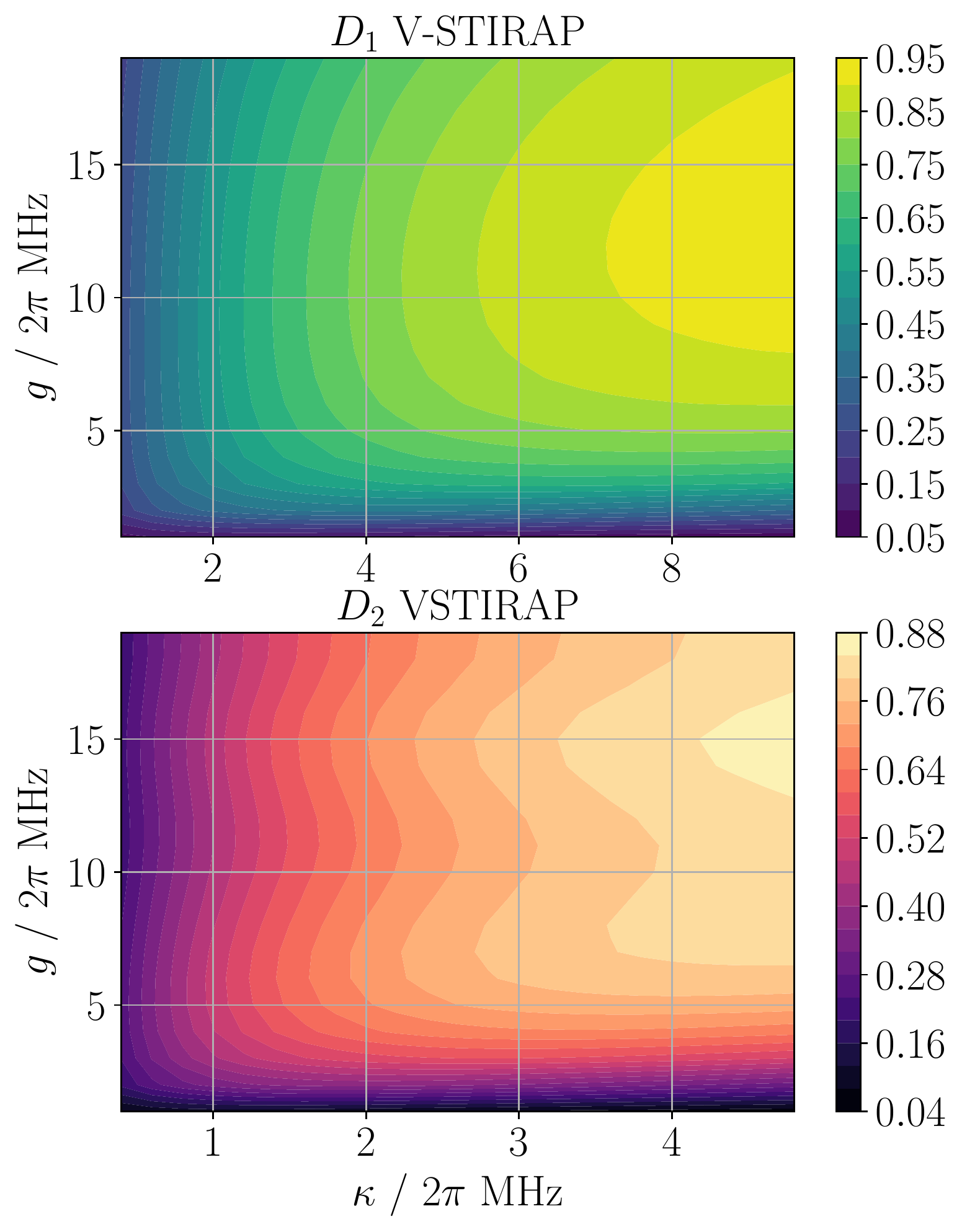}
    \caption{V-STIRAP Photon Production efficiency for both $D_1$ and $D_2$ lines plotted for various atom cavity coupling strengths $g$ and photonic field decay rates $\kappa$. The laser pulse length is fixed at $500$ ns and it takes a $\sin^2$ shape. Each point in the plot corresponds to a local optimisation problem to find the optimal detuning and peak Rabi frequency to maximise photon production efficiency. For the same photonic field decay rate $\kappa$ and for the same atom cavity coupling $g$ the $D_1$ scheme is more efficient than the $D_2$ scheme. Note, however, that $g$ for different atomic transitions does not correspond to the same physical mirror parameters, since the Clebsch-Gordan coefficients for the V-STIRAP transition differ.}
    \label{fig:vstirap_contour_d1d2}
\end{figure}
\begin{figure}[ht]
    \centering
    \includegraphics[width=\columnwidth]{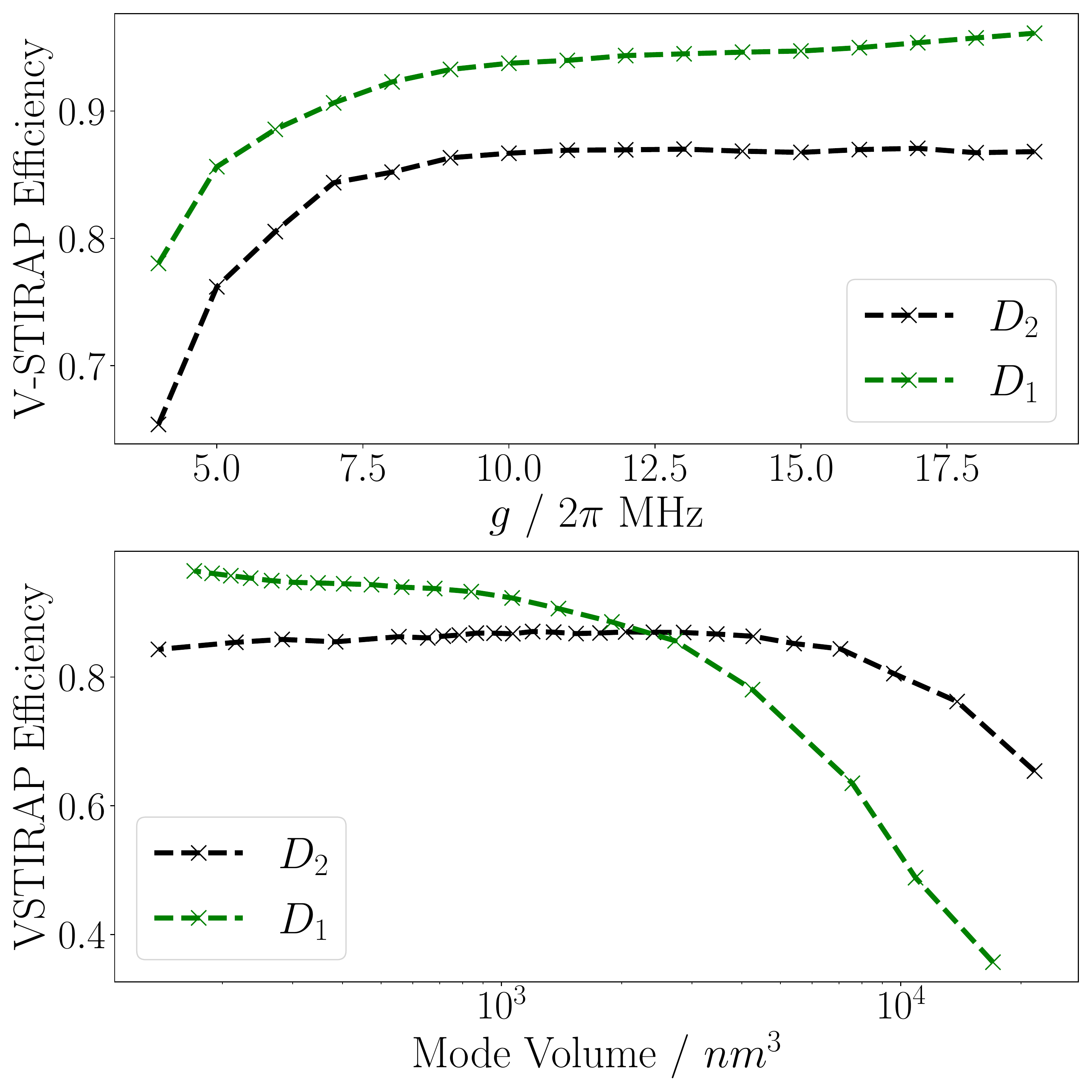}
    \caption{Efficiency of the V-STIRAP photon production procedure for various atom cavity coupling strengths $g$ for both the $D_1$ and $D_2$ lines (top), at fixed $\kappa$. Given the desire to reasonably compare physical cavity parameters across possible schemes the efficiency is also plotted against mode volume which is a function of the cavity mirror properties (bottom). It is clearly visible, that for most feasible cavity mode volumes $> 2 \times 10^3 \mathrm{nm}^3$, the $D_2$ V-STIRAP process is more efficient than the $D_1$ V-STIRAP process. }
    \label{fig:vst-real-atom-eff}
    
\end{figure}
The V-STIRAP process lies at the heart of the photon production scheme and its merits have been demonstrated for an idealised three level system in Sec. \ref{sec:vst}. We now consider the full atomic structure of $^{87}\mathrm{Rb}$ and carefully determine the optimal parameters and efficiency trade-offs exemplified by this real atomic quantum system. \\
The relevant atomic states are shown in Fig \ref{fig:exp-sequence}. After initialising the atom into the $\ket{F=2, m_F=0}$ ground state , a $\pi$ polarised driving pulse, near resonant with the transition to $\ket{F'=1, m_{F'}=0}$ addresses the atom and the cavity is tuned to be resonant, such that one transfers the atom into the superposition state:
\begin{equation}
\label{eq:rb_superposition}
    \ket{\Phi_{\phi=\pi}}=\frac{1}{\sqrt{2}}(\ket{F=1, m_F=1}-\ket{F=1, m_{F}=-1}).
\end{equation} A linearly polarised photon is then emitted preferentially through one of the cavity mirrors. 

It was shown in Sec. \ref{sec:vst} that in order to push the efficiencies near unity, one requires strong atom cavity coupling with respect to the atomic depolarisation rate and long photons or, alternatively, a fast photonic field decay. Some specific cavity regimes are explored in Fig. \ref{fig:vstirap_contour_d1d2}.  We consider a fixed pulse of length $500$ns which takes the following form: $\Omega_{\mathrm{VST}}=\Omega_{\mathrm{max}}\sin^2(\pi t/T)$. Every particular cavity parameter configuration corresponds to a local optimisation problem with regards to the detuning $\Delta$ and peak Rabi frequency of the laser pulse $\Omega_{\mathrm{VST}}$. These are numerically optimised for maximising the photon production probability (cf. Eq. \ref{eq:eff_1}). In general, the optimal peak Rabi frequency increases with regards to an increase in atom-cavity coupling (cf. App. \ref{sec:app_rb_vst} for further details). \\
Fig. \ref{fig:vstirap_contour_d1d2} shows that the $D_1$ transitions lead to higher efficiencies than the $D_2$ transitions. To compare the two possible transition lines more explicitly, we consider a realistic photonic field decay rate of $\kappa=2 \times 2\pi$ MHz and vary only the atom-cavity coupling strength $g$ as exemplified in Fig. \ref{fig:vst-real-atom-eff}.  Importantly, unlike in an idealised three level system (cf. Fig. \ref{fig:vst_ideal_length}), the efficiency does not asymptotically approach a theoretical maximum, as the atom cavity coupling strength (and cooperativity) further increase (cf. Fig. \ref{fig:vst-real-atom-eff}). This can be attributed to the fact that, stronger coupling necessitates higher peak Rabi frequencies $\Omega_{\mathrm{VST}}$ which lead to undesirable off-resonant couplings to other excited levels. For the $D_2$ line, the excited levels $F'=1$ and $F'=2$ are split only by $\approx 160$ MHz \cite{steck87}. As the strength of the atom-cavity coupling increases beyond $\approx 10 \times 2\pi$ MHz, we observe no further increase in V-STIRAP photon production efficiency on $D_2$ as shown in Fig. \ref{fig:vst-real-atom-eff}. The excited levels $F'=1,2$ for the $D_1$ transitions are significantly further detuned at $\approx 816$ MHz. Thus, for the $D_1$ scheme, a decline in efficiency is not visible up to the couplings shown in Fig. \ref{fig:vst-real-atom-eff}.

It shall also be stressed that the same physical cavity mirror parameters do not give rise to the same atom-cavity coupling strength for different atomic transitions, because the atom cavity coupling $g$ is a function of the Clebsch-Gordan coefficients of the particular transitions between the states $\ket{g_1^{\pm}}, \ket{x}$ and $\ket{g_2}$. For the same physical cavity parameters, the coupling strengths scale as $g_{D_2}\approx g_{D_1} \times 2.3$. The trade-off between cavity mode volume and single photon production efficiency is also shown in Fig. \ref{fig:vst-real-atom-eff} to compare existing cavity designs. There exists a cross-over point at a mode volume of around $2 \times 10^3 \mathrm{nm}^3 $ below which it becomes advantageous to use the $D_1$ line for photon production, but for most existing physical cavities \cite{micro-cav_doherty}, the $D_2$ line is preferential.

\subsection{STIRAP Re-preparation}
\begin{figure}[ht]
    \centering
    \includegraphics[width=\columnwidth]{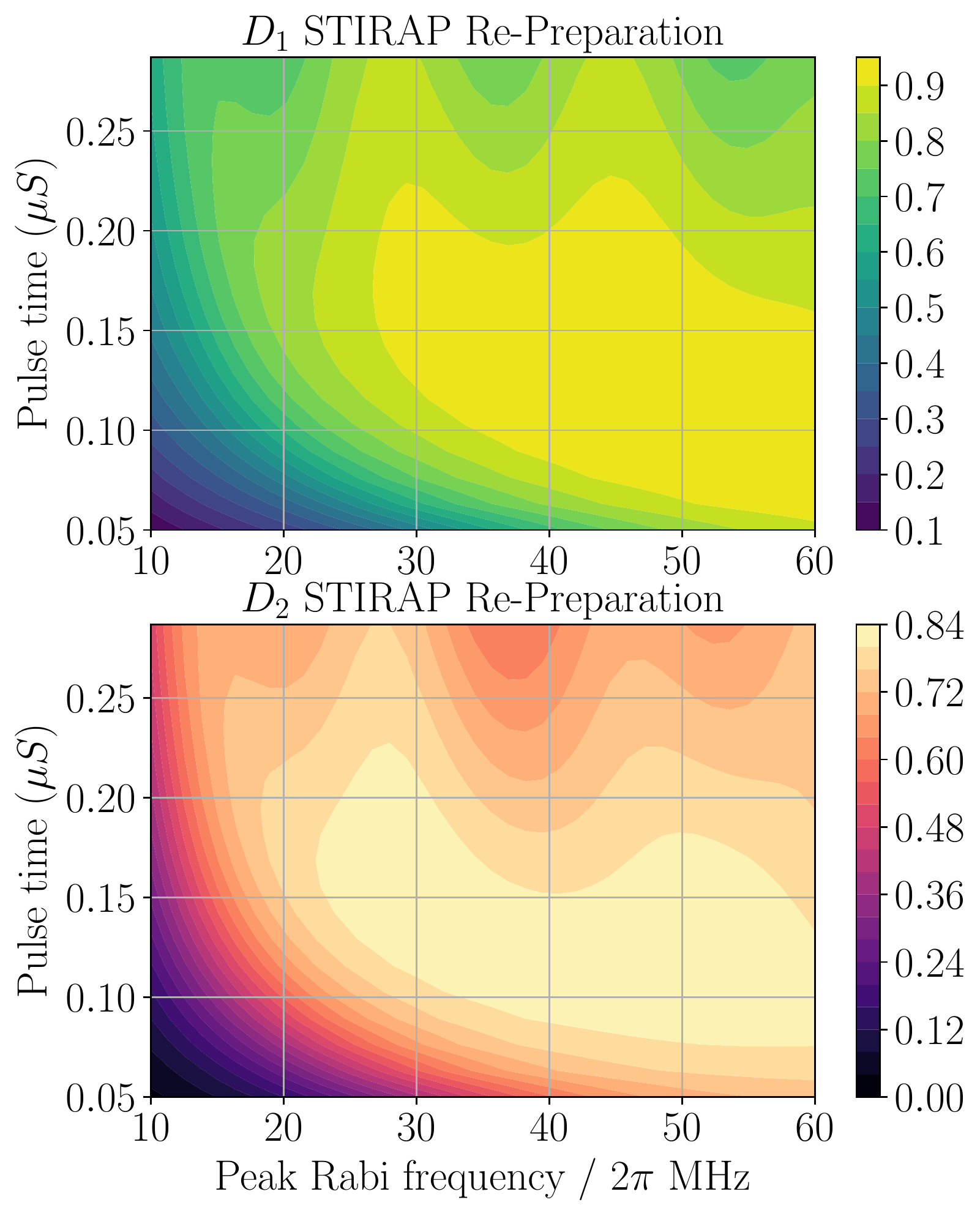}
    \caption{STIRAP re-preparation efficiency for both the $D_1$ and $D_2$ transitions plotted against the length of the re-preparation pulses for various peak Rabi frequencies $\Omega$. STIRAP re-preparation is generally more efficient on $D_1$ and, unlike in an idealised system, arbitrarily large peak Rabi frequencies do not further increase the re-preparation efficiency as undesired off-resonance effects adversely affect the re-preparation efficiency.}
    \label{fig:repumping_contour_d1d2}
\end{figure}
The coherent re-preparation scheme described in Sec. \ref{sec:ideal_cr} can be applied to the real atomic system of $^{87}\mathrm{Rb}$, which offers either the $D_1$ or $D_2$ transitions for coherently re-preparing the atom back into the desired initial state with two counter-intuitively sequenced laser pulses of appropriate polarisations (cf. App. \ref{sec:app_rb_rep} for further details). In general, when simulating the full atomic structure we observe similar trade-offs noted in the previous section for polarised photon production with V-STIRAP.\\
We follow the same techniques introduced in Sec. \ref{sec:ideal_cr} and consider two interleaved pulses of constant rms Rabi frequency with a hyper-Gaussian mask \cite{vitanov_2009} with numerically optimised shape and peak Rabi frequency (cf. App. \ref{sec:app_rb_rep} for further details). This leads to the highest population transfer efficiencies at the desired time-scales. As for the photon production process, off-resonant coupling to other excited levels is the main source of depolarisation and adversely affects the repumping efficiency, particularly for extremely fast repumping which requires high peak Rabi frequencies, as exemplified in Fig. \ref{fig:repumping_contour_d1d2} for both the $D_1$ and $D_2$ transitions. We observe that for re-preparation times of up to $150 \text{ ns}$ the maximal population transfer efficiency for re-preparation on these lines is $0.95$ and $0.83$ respectively. Like in the V-STIRAP process, the relative proximity of the higher lying excited levels detrimentally affects the efficiency of the re-preparation with respect to the $D_2$ transitions. 
It should be noted, that the presence of the cavity may also adversely affect the re-preparation efficiency, as it produces photons on the same transitions used to coherently re-prepare the atom back into the initial state. However, significantly detuning the cavity from the resonantly driven transition to $F'=1, m_F=0$ ensures that there is no such effect. \\
Due to the reduced achievable efficiencies with the $D_2$ line (cf. Fig. \ref{fig:repumping_contour_d1d2}) and the adverse affect of a near-resonant cavity with respect to the re-preparation efficiency, we conclude that, given current state of the art cavity parameters \cite{micro-cav_doherty}, the optimal photon production scheme uses a combination of V-STIRAP with respect to the $D_2$ transitions for single photon production, as well as STIRAP re-preparation via the $D_1$ transitions. It exploits the optimal efficiency of the underlying processes and avoids the introduction of further losses in efficiency, as the cavity is now extremely far detuned from the transitions used for quickly re-preparing the atom back into the desired initial state. 
\subsection{Optical Pumping}
\label{sec:op}
 An  $^{87}\mathrm{Rb}$ atom loaded into an optical cavity is initially populated in a random ground state. Moreover, during the photon production cycle consisting of V-STIRAP and coherent re-preparation, the atom may spontaneously decay into any Zeeman state. For these reasons, it is necessary to use an optical pumping process to efficiently prepare the atom into the desired initial $\ket{F=2, m_F=0}$ state.

\begin{figure}[ht]
    \centering
    \includegraphics[width=\columnwidth]{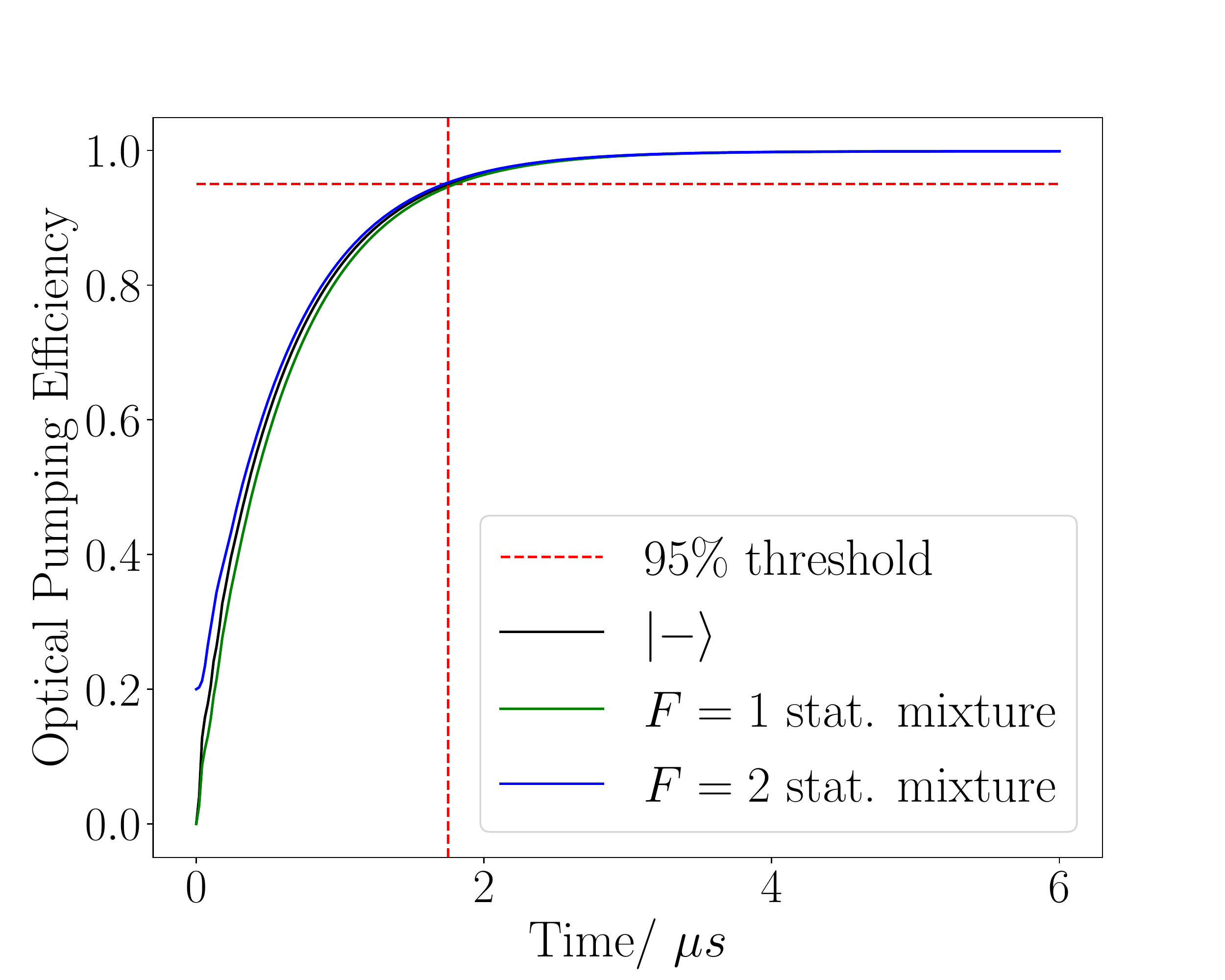}
    \caption{Optical Pumping Efficiency plotted as a function of time for various initial conditions for the $D_1$ scheme (cf. App. \ref{sec:app_rb_op} for $D_2$). The black trace corresponds to an initial state corresponding to the coherent superposition of the $F=1$ groundsates $\ket{\Phi_{\pi}}$(cf. Eq. \ref{eq:rb_superposition}), the green and blue traces correspond to statistical mixtures of the states of the $F=1$ and $F=2$ levels respectively. Re-preparation efficiency depends on the cost one is willing to incur with respect to the repetition rate. For instance, after $ \approx 1.75 \mu$s the atom is repumped to the desired state with a likelihood of 95 \% (cf. dotted red line).}
    \label{fig:op_init}
    
\end{figure}

Optical pumping for the $D_1$ and $D_2$ transitions requires two $\pi$ polarised lasers (with respect to the $y$ axis, travelling along the $x$ axis, as shown in Fig. \ref{fig:exp-sequence}) which are nearly resonant with the $F=1 \rightarrow F'=2$ and $F=2 \rightarrow F'=2$ transitions. Both lasers act simultaneously and through spontaneous emission all population eventually accumulates in the $F=2, m_F=0$ groundstate, because the transition from $F=2, m_F=0$ $\rightarrow$ $F'=2, m_F'=0$ is dipole forbidden. This is shown in Fig \ref{fig:exp-sequence}. \\ 
Resonant pulses exhibit low population transfer efficiencies due to the creation of two-photon resonances which would repopulate other states of the $F = 2$ level, due to the formation of unwanted dark states formed by superpositions of individual sub states of $F = 1$ and $F = 2$. The detunings of both $\pi$ polarised pulses from resonance, as well as their peak Rabi frequencies are numerically optimised for maximal population transfer to the desired state $\ket{F=2, m_F=0}$, cf. App.\ref{sec:app_rb_op} for a detailed description of the parameters.  The efficiency with which initialisation or re-preparation with optical pumping occurs over different time scales and various initial conditions is shown in Fig. \ref{fig:op_init} for the $D_1$ line. This is more efficient than using the $D_2$ transitions, which has additional excited atomic levels that adversely affect the population transfer efficiency. In principle, one can achieve arbitrarily high efficiencies with extremely long pulses. It should be stressed that the time scale at which optical pumping occurs is about an order of magnitude longer than that of coherent re-preparation because of the random walk of the population back into the desired initial state.

\subsection{Discussion}
\label{sec:discussion}
Following the analysis of the previous two sections, we consider the combination of optical pumping, V-STIRAP, and STIRAP with respect to the $D_2$ and $D_1$ transitions, respectively, for polarised photon stream production. 
\begin{figure}[ht]
   \centering
    \includegraphics[width=\columnwidth]{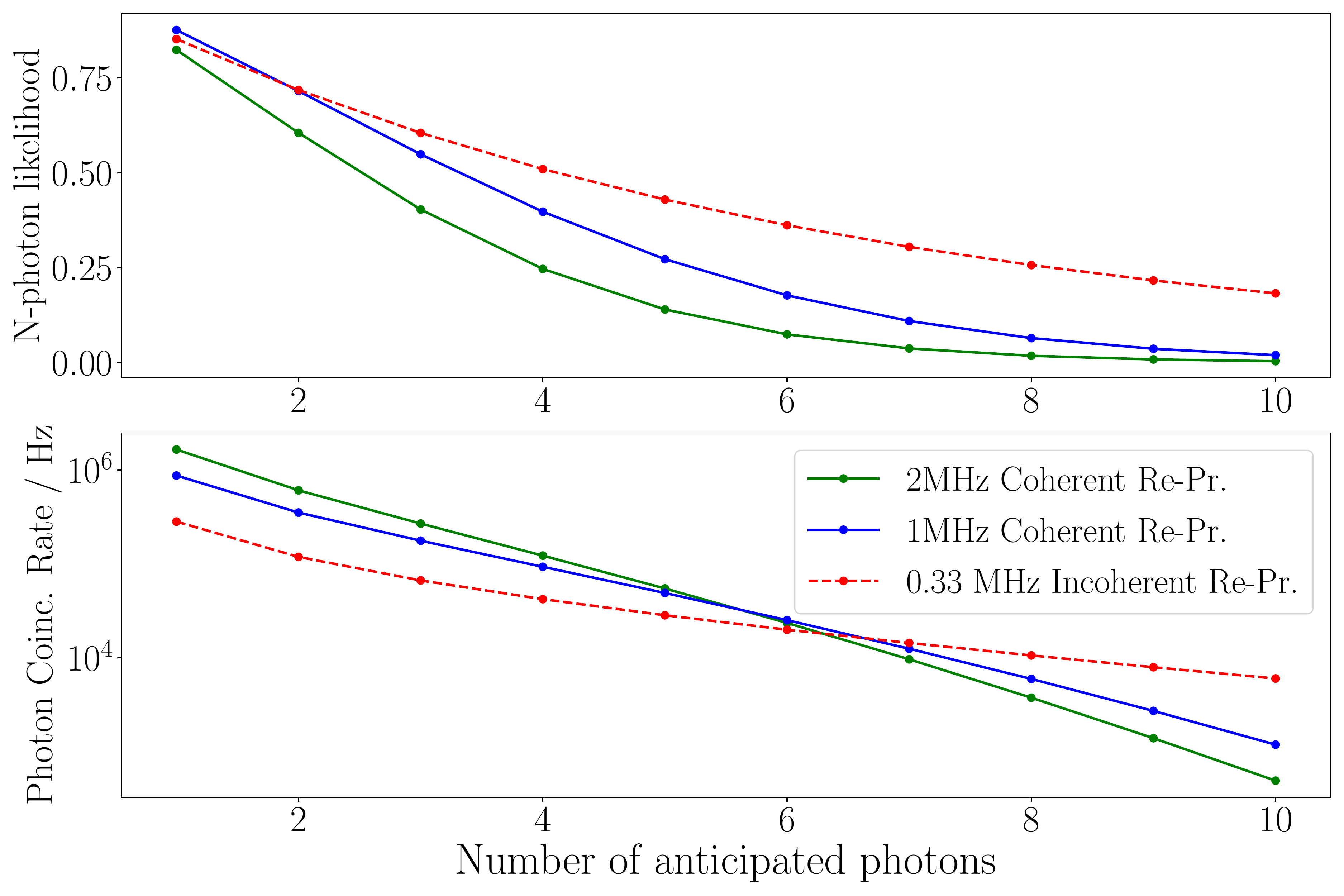}
    \caption{Polarised photon stream production likelihood (top) and expected n-photon sequence coincidence rate ($\text{Efficiency} \times \text{Single Photon Repetition Rate} / \mathrm{N}$), where N denotes the number of photons (bottom) for existing cavity design \cite{micro-cav_doherty}. Slower driving procedures with longer photons or incoherent re-preparation methods lead to higher coincidence counts for longer photon sequences as the efficiencies are higher and thus the N-photon production efficiencies scale more favourably.}
    \label{fig:eff_sequence}
\end{figure}
We present simulated efficiencies for generating bursts of polarised photons with recurring cycles of $D_2$ V-STIRAP photon production and $D_1$ STIRAP re-preparation from a state of the art optical cavity \cite{micro-cav_doherty} and $^{87}\mathrm{Rb}$ atoms - which for simplicity are assumed to be stationary \footnote{For the sake of simplicity, atoms in the simulation are assumed to be stationary. In the case of atoms in free flight or trapped in a conservative trap, we note that each VSTIRAP/STIRAP cycle imparts a maximal momentum change of $3 \hbar k$ on each atom (in reality this will be much lower), thus increasing the temperature of a trapped Rb atom by approximately 538nK and leading to a corresponding Doppler shift of approximately 22.48 kHz. This might reduce the efficiency of the process if the burst length is of the order of hundreds of photons, but it is negligible for the photon burst lengths described in the paper.} - in Fig. \ref{fig:eff_sequence}. It should be noted that the proposal outlined in Sec. \ref{sec:ideal} and applied to $^{87}\mathrm{Rb}$ describes a perfectly degenerate state configuration, yet perfect degeneracy is nearly impossible to achieve due to the small B-field fluctuations. If these cause Larmor precessions on the time scales of the photons, indistinguishability is affected as the polarisation is no longer constant in time. One can mitigate this by imposing a small constant external field along the preferred quantisation axis (i.e. $y$) which induces a hyperfine state splitting smaller than the line-width of the transition or the cavity. In this case, the scheme is robust in the face of small external magnetic fields and a detailed technical discussion is provided in App. \ref{sec:app_quant-rb}. \\ \\

The repetition rate of the V-STIRAP and STIRAP sequence can be altered by modifiying the duration of either or both. In what follows we fix the coherent STIRAP re-preparation duration to 140ns.  Shorter re-preparation pulses lead to reduced efficiencies without significantly improving the overall photon production repetition rate and longer pulses only marginally increase efficiency. We do however consider two different V-STIRAP photon driving pulses of length $360$ and $860$ ns, which lead to overall repetition rates of $2$ and $1$ MHz respectively. Longer driving leads to higher single photon production efficiencies, as evident in Fig. \ref{fig:eff_sequence}, yet the expected number of single photon counts in a particular time interval is lower. However, when the photon burst length exceeds six or more consecutive photons, the expected count rate is larger for the photons which are driven more slowly at a repetition rate of $1$ MHz.

Neither the V-STIRAP photon production, nor the STIRAP re-preparation are perfectly efficient. Thus, it is important to characterise the effects of decoherence on the produced photon stream. Atomic depolarisation during the V-STIRAP procedure may manifest itself, such that the atom is not prepared into a superposition of the states of the $F=1$ level. This adversely affects the coherent STIRAP re-preparation efficiency which only correctly re-prepares the atom into the initial state if starting from the superposition state $\ket{\Phi_{\pi}}$. Moreoever, atomic depolarisation during STIRAP re-preparation can similarly lead to incoherent decay of the atom into the wrong Zeeman state.
Simulations for a state of the art optical cavity \cite{micro-cav_doherty} and optimal pulse parameters show that for 10 consecutive photon production attempts with STIRAP coherent re-pumping, a non-negligible accrual of population in other states of the $F=2$ manifold is unavoidable. The population of undesired Zeeman states however leads to a small, efficiency gain with respect to the production of photon sequences with fixed linear polarisation. Even when the atomic population falls to the wrong Zeeman state after re-preparation, we can obtain a useful photon (of desired $H$ polarisation). This is exemplified in Tab. \ref{tab:pol_repump} (for further details see App. \ref{sec:app_full-scheme}), as for instance when starting in the state $F=2, m_F=-2$ the probability of emitting an H-polarised photon is $0.188$. \\
Practically, undesired $\pi$ polarised photons would be detected at the cavity output with a polarising beam splitter, indicating the failure of a particular coherent photon productions sequence. This could be used to immediately trigger incoherent re-preparation with optical pumping before initialising another coherent photon production sequence.\\

We also consider a possible sequence where the atom is re-prepared entirely incoherently with optical pumping lasting for 2.5 $\mu$s and with the V-STIRAP process lasting for 500ns, yielding a single photon repetition rate of $0.33$MHz. This has the advantage of achieving repumping efficiencies close to unity for longer time scales, regardless of the initial atomic state after photon production. This of course incurs other costs; the repetition rate is significantly reduced and the atom does not maintain coherence throughout the process which precludes the performance of quantum information processing tasks with the atom as an entanglement mediator (cf. \cite{rempe_cluster}). However, the likelihood to obtain counts for longer photon sequences ($>7$) exceeds those achievable with coherent re-preparation.

\begin{table}

\begin{tabular}{l|l|l|l|l|l}

\hline
$F=2, m_F = $ & -2    & -1    & 0     & 1     & 2     \\ \hline
$p(H)$       & 0.188     & 0.158  & 0.853 & 0.158  & 0.188     \\ \hline
$p(\pi)$        & 0.049 & 0.710 & 0.009  & 0.710 & 0.049 \\ \hline
Purity        & 0.8 & 0.182 & 0.99  & 0.182 & 0.8 \\ \hline
\end{tabular}
\caption{Probability of emitting a photon of particular polarisation, for a cavity with $C \approx 10$ \cite{micro-cav_doherty} when applying the optimised driving pulse when correctly initialised (i.e. $m_F=0$) and when incorrectly initialised into the other sub-states. The purity, i.e. $p(H) / (p(H)+p(\pi))$ is also shown and decreases significantly for the $m_{F}=\pm 1$ sub-states.}
\label{tab:pol_repump}
\end{table}

\section{Outlook} \label{sec:outlook}
We have developed a novel scheme for polarisation controlled single-photon emission from atom-cavity systems, and have underpinned with extensive numerical simulations that it might be used for the production of $n$-photon bursts in a deterministic manner. Combined with coherent re-preparation, expected success and coincidence rates are unprecedented. The simulated single photon production efficiency with a state of the art optical cavity \cite{micro-cav_doherty} reaches $85 \%$ which to the knowledge of the authors lies beyond any existing single photon production efficiencies achieved to date with atoms at a comparable repetition rate (similar efficiencies are demonstrated with ions \cite{Barros_2009}, yet at a repetition rate several order of magnitudes lower). Moreover, improved cavity designs with higher atom-cavity coupling (e.g. fibre tip cavities cf. \cite{Hunger_2010}) would allow for single photon production efficiencies $>90 \%$. It should however be noted that, even with increased atom-cavity coupling strength, the extremely high peak Rabi frequencies would not also induce off-resonant coupling errors, but would also induce differential Stark shifts which could be compensated with frequency chirping.

Compared to non-deterministic physical platforms like SPDC sources \cite{spdc_bosonsampling_pan}, atom-cavity sources do not reach comparable repetition rates. Nevertheless, our proposal demonstrates the ability to efficiently generate highly indistinguishable photons at the MHz scale. 
The high single photon production efficiencies of deterministic sources, lead to a favourable scaling behaviour for larger photon sequences. The use of a single photon source for n-photon burst production is limited to a scaling of $p_1^n$ at best, where $p_1$ represents the single photon production efficiency and $n$ represents the photon burst length. Consequently, when dealing with non-deterministic sources, which frequently exhibit low $p_1$, the probability of larger n-photon burst production rapidly decreases. Besides atomic sources, other deterministic single photon sources offer similar advantages. Recently, much progress has been made in demonstrating efficient and fast photon production with quantum dots \cite{bright_fast_sps, qudot-pan, Somaschi2016}.

Theoretically, we have shown 10 consecutive photon streams with a production efficiency of up to around $2 \%$ and $20 \%$ respectively for coherent and incoherent re-preparation methods, as well as expected $10$ photon counts rates on order of kHz which represents a significant improvement from any previous atom cavity scheme. This is essential for demonstrating multi-photon coincidences with high clockrates \cite{Munzberg2022} for boson sampling \cite{XanaduBosonSampling2022} or other quantum information processing applications. Moreover, preserving coherence of the emitter throughout the entire process is another extremely appealing feature. It has been demonstrated in \cite{rempe_cluster} that for optical GHZ and cluster state generation with more than 10 photons, atoms significantly outperform quantum dots \cite{schwarz_qud_cluster, pan_qud_cluster} or state of the art spontaneous parametric down conversion sources \cite{spdc_bosonsampling_pan}. Creating large highly entangled optical states is instrumental for the realisation of measurement based quantum computing \cite{mbqc}. 
In addition to the extensive exploration of $^{87}\text{Rb}$, it would be interesting to consider applying the scheme to other neutral atoms (e.g. Caesium \cite{cesium_kimble}) or ions (e.g. Calcium \cite{Barros_2009}, Strontium \cite{strontium}), which exhibit similar level structures.

\begin{acknowledgments}

The authors thank Thomas Doherty, Reuben Sorsbie, Mark IJspeert and Chloe So for useful discussions.
J.O.E. acknowledges support from the Studienstiftung des Deutschen Volkes. J.R.A. acknowledges funding by the European Union Horizon 2020 (Marie Sklodowska-Curie 765075-LIMQUET). This work was supported by the EPSRC through the quantum technologies programme (NQIT hub, EP/M013243/1).

\end{acknowledgments}

\bibliography{refs}

\newpage

\appendix
\renewcommand\thefigure{\thesection.\arabic{figure}}  
\renewcommand\thetable{\thesection.\arabic{table}} 
\setcounter{figure}{0}

\section{Efficiency in idealised system}
\label{sec:app_eff}
  In the idealised configuration, the desired coherence between $\ket{g_1^+} \text{and} \ket{g_1^-}$ can only be attained from the initial state $\ket{g_2}$ by adiabatically transferring population through the dark state with the coherent system dynamics. Spontaneous emission may populate a statistical mixture of the two ground-states $\ket{g_1^+} \text{and} \ket{g_1^-}$, yet an incoherent process cannot lead to the creation of coherence between the ground states when the system is initialised in $\ket{g_2}$. Thus, we can also estimate the efficiency as: 
 \begin{equation}
 \label{eq:eff_2}
     \eta_{2}= \bra{\Phi_{\pi},0} \rho(t_f) \ket{\Phi_{\pi},0}- \bra{\Phi_{0},0} \rho(t_f) \ket{\Phi_{0},0}\underset{\textit{ideal sys.}}{=} \eta_{1},
 \end{equation}
 where $\rho(t_f)$ denotes the density matrix at the end of the simulation and $\ket{\Phi,n}$ represents a tensor product of the atomic and Fock state, with $\Phi_{\phi}=\frac{1}{\sqrt{2}}\left(\ket{g_1^+}+e^{i\phi}\ket{g_1^-} \right)$.
\section{Simulation Parameters}
The simulations for this paper were performed with the QuTiP \cite{JOHANSSON20121760} package in Python. The evolution of the states under coherent laser dynamics and incoherent cavity and spontaneous decay dynamics are determined by numerically solving the Lindblad master equation for various physical parameters and atomic energy level configurations. The code is available on \href{https://github.com/jan-o-e/rb_photon_prod}{GitHub}.
The following section aims to give a more detailed overview of the details and explicit parameters used for the results and plots produced in the main paper.

\label{sec:sim-params}
\subsection{Ideal three level system}
\subsubsection{V-STIRAP Photon Production}
\label{sec:app_ideal_vst}
\begin{figure}[ht]
    \centering
    \includegraphics[width=\columnwidth]{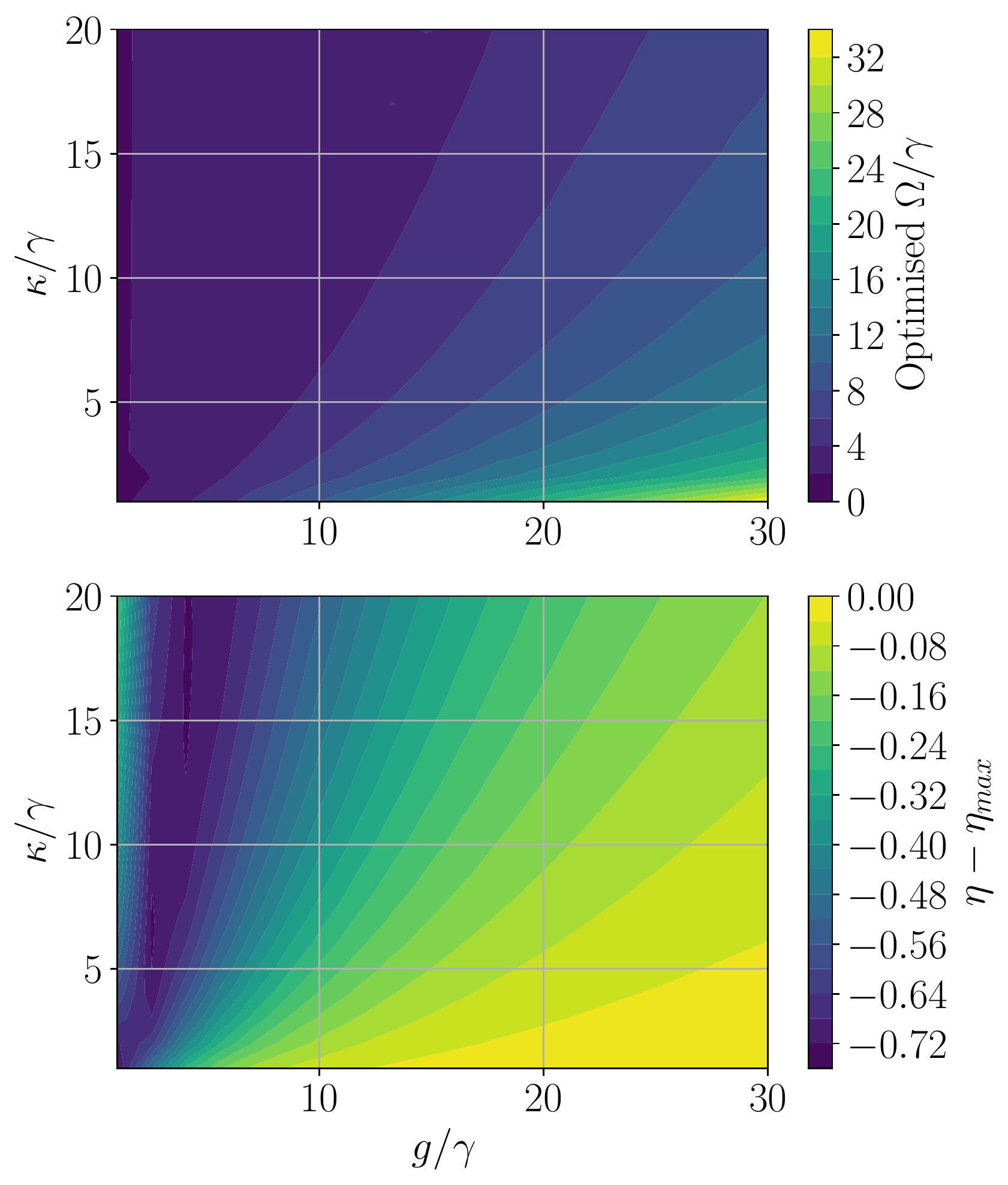}
    \caption{Optimised peak Rabi frequency for V-STIRAP photon production in idealised three level system, at the corresponding data points $g / \gamma, \kappa / \gamma$) (top) and difference from the theoretical upper limit in efficiency $\eta_{\mathrm{max}}=2C / (2C+1)$, where $C$ is the cavity cooperativity (bottom). }
    \label{fig:ideal_vst_omega}
\end{figure}
In Sec. \ref{sec:ideal} an idealised atomic emitter was considered with four configured states ${\ket{x}, \ket{g_2}, \ket{g_1^-}, \text{and} \ket{g_1^+}}$ and various cavity parameters for producing photons. The V-STIRAP driving pulse took the following form: $\Omega_{\mathrm{VST}}(t) = \Omega_{\mathrm{max}} \sin^2(\pi t / T )$, where $\Omega_{\mathrm{max}}$ is the peak Rabi frequency and $T=10\gamma$ (different pulse lengths were omitted in favour of brevity and the trends for different pulse lengths were generally similar). The peak Rabi frequencies were numerically optimised for maximising photon production probability. Since the system is idealised, we normalised everything with respect to $\gamma$ and then varied $g$ and $\kappa$, the atom cavity coupling, as well as the photonic field decay rate $\kappa$. As evident in Fig. \ref{fig:ideal_vst_omega}, the optimal peak Rabi frequencies for these atom-cavity coupling strengths increase drastically as a function of $g$, especially for fast field decay. Changing the detuning from the state $\ket{x}$ has close to no effect on the efficiency calculation and hence all V-STIRAP processes considered here are resonant. It should also be noted that we only approach the theoretical upper limit in photon production efficiency (cf. Fig. \ref{sec:app_ideal_vst}) for higher cavity cooperativities in the regimes of high $g$ and low $\kappa$ or for very large $\kappa$, i.e. in the fast cavity regime.\\

\subsubsection{STIRAP Re-preparation}
\label{sec:app_ideal_rep}
\begin{figure}[ht]
    \centering
    \includegraphics[width=\columnwidth]{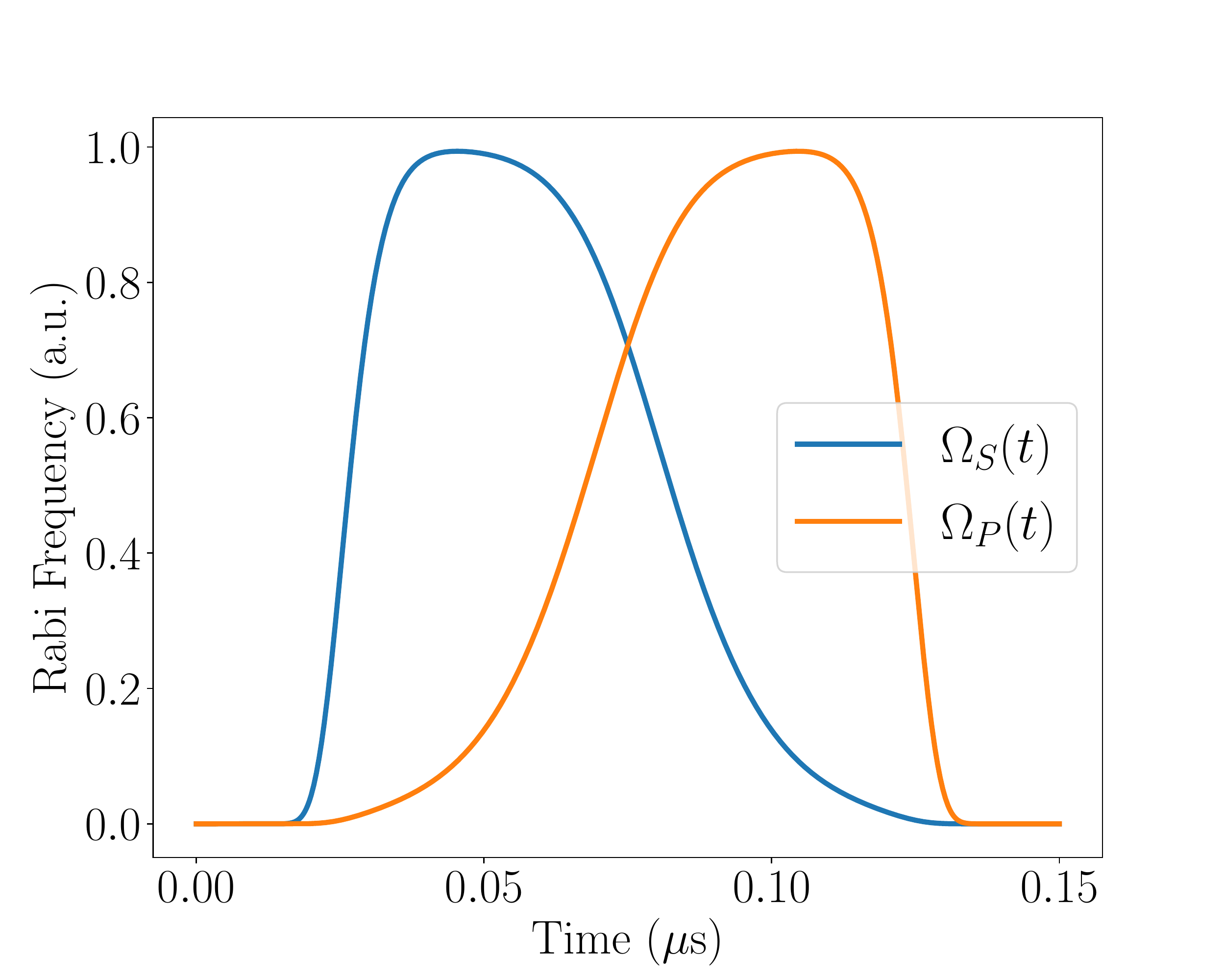}
    \caption{STIRAP re-preparation pulse shapes generated by optimising the shape of the Gaussian mask for maximal population transfer efficiency in an idealised atomic system.}
    \label{fig:shaped-repumping}
\end{figure}

We follow a technique introduced in \cite{vitanov_2009} and consider two pulses of constant rms Rabi frequency $\sqrt{(\Omega_{S}^2+\Omega_{P}^2)}$ with a hyper-Gaussian mask with numerically optimised shape. The analytic expressions for the pulse shapes of total duration $T$ and peak Rabi frequency $\Omega_{\mathrm{max}}$ are as follows:

\begin{equation}
\begin{aligned}\Omega_{S}(t)=\Omega_{\mathrm{max}}\cdot e^{\left(-\left(\frac{t-T/2}{c}\right)^{2n}\right)}\sin\left(\frac{\pi}{2\cdot\left(1+e^{-\left(\frac{a\left(t-T/2\right)}{T}\right)}\right)}\right)\\
\Omega_{P}(t)=\Omega_{\mathrm{max}}\cdot e^{\left(-\left(\frac{t-T/2}{c}\right)^{2n}\right)}\cos\left(\frac{\pi}{2\cdot\left(1+e^{-\left(\frac{a\left(t-T/2\right)}{T}\right)}\right)}\right)
\end{aligned}
\label{eq:masked_pulses}
\end{equation}
where $(n,a)$ are control parameters and $c=0.05$ is chosen such that both pulses exhibit (within the numerical limit of the solver) zero amplitude at $t=0$ and $t=T$. This is a desiderata for making the simulation more realistic. The optimal values of $(n,a)$ for the idealised system are $(6,14)$. The pulses exhibited shown in Fig. \ref{fig:shaped-repumping} lead to re-preparation efficiencies which approach unity at large peak Rabi frequencies. However, this method does not take into account the entire structure of a real atom where further losses arise from off-resonant coupling to undesired levels.

\subsection{Real Atom: Rubidium 87}
In Sec. \ref{sec:ph_prod} we consider a real Rubidium atom with all the configured energy levels and their hyperfine states for both the $D_1$ and $D_2$ lines respectively. For $D_1$ this encompasses $F=(1,2)$ and $F'=(1,2)$ levels, such that there are 8 ground and excited states respectively. For $D_2$ this requires the addition of $F'=0$ and $F'=3$ levels such that there are a total of 16 excited states. The couplings and state configuration used in all subsequent simulation, are for the perfectly degenerate case (excluding Sec. \ref{sec:app_quant-rb} where we explicitly vary the external magnetic field). \\
\subsubsection{Quantisation Axis \& External Magnetic Field}
\label{sec:app_quant-rb}
\begin{figure}[ht]
   \centering
    \includegraphics[width=\columnwidth]{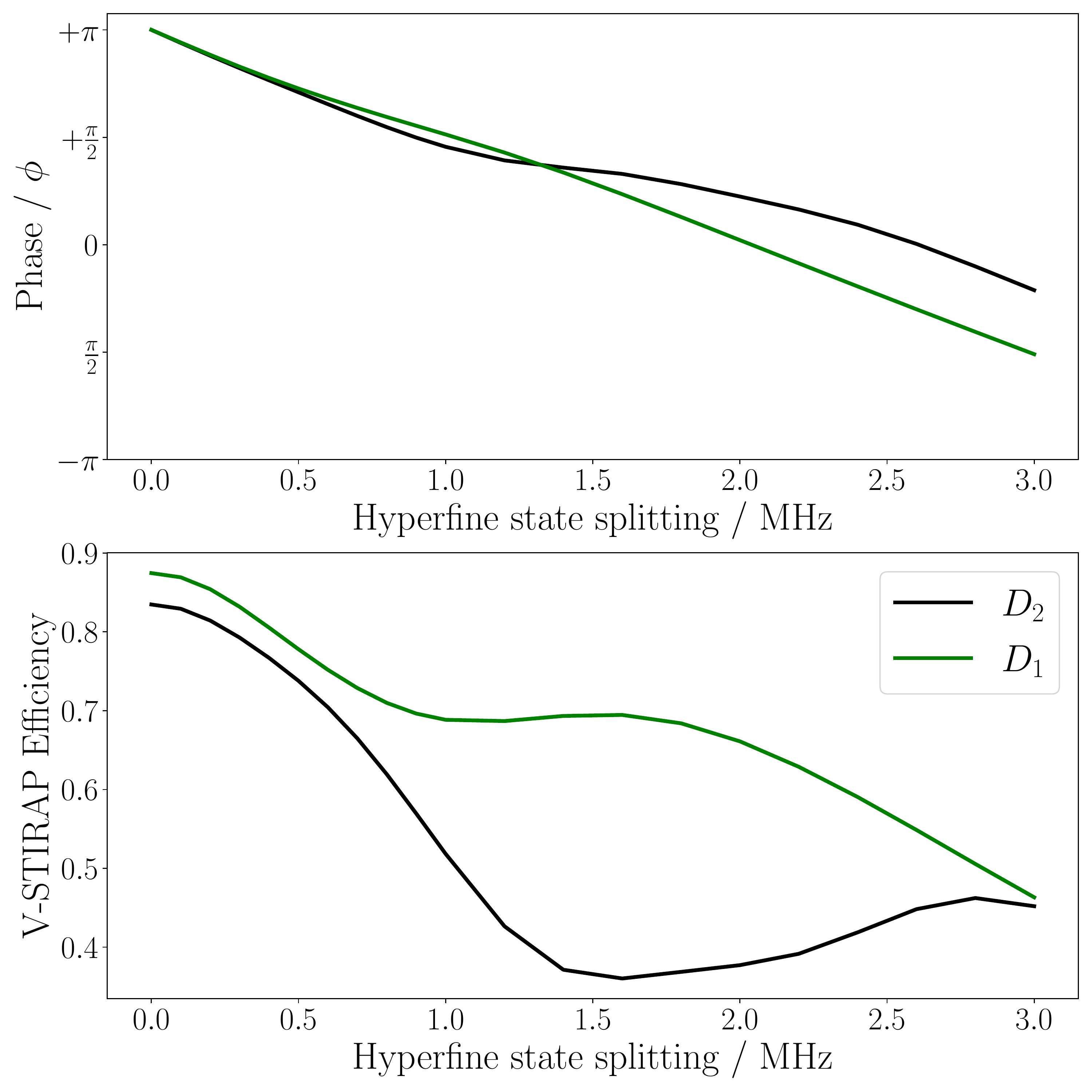}
    \caption{Variation of the phase $\phi$ between the two hyperfine groundstate $F=1, m_F=\pm 1$ and hyperfine state splitting by imposing a weak external magnetic field aligned with the preferred quantisation axis (top) and variation of the V-STIRAP photon production efficiencies for state of the art optical cavity \cite{micro-cav_doherty} and photon of length $500$ ns (bottom). The phase rotates linearly for $D_1$, yet varies slightly differently for $D_2$ from about $1$ MHz splitting onwards, as the $D_2$ transitions experience an asymmetric variation in the dipole matrix elements for the $\sigma^{\pm}$ transitions which couple to the individual states in \ref{eq:phase}. This is also reflected in the sharp decline in efficiency for the $D_2$ transitions, which eventually increases again at larger splittings. A field of $1$ G corresponds to a splitting of $0.7$ MHz and in reality we seek to remain in the mG regime, such that there is close to no decline in efficiency from the perfectly degenerate state configuration. }

    \label{fig:bfield-split}
\end{figure}
The proposed photon production scheme does not require a strong external magnetic field to significantly split the hyperfine states, such that they can be addressed individually. As shown in \cite{barrettNonlinearZeemanEffects2018}, in $^{87}\mathrm{Rb}$ this approach led to the apparition of non-linear Zeeman effects which strongly curtails achievable efficiencies. To ensure that the atomic quantisation does not vary erratically in time, we do however seek to impose a weak external magnetic field, otherwise background field fluctuations would adversely affect photon polarisation purity within a particular burst. The achievable simulated efficiencies in Sec. \ref{sec:rb87} all assume perfectly degenerate ground-states, yet in reality we seek to impose a small external magnetic field, aligned with the quantisation axis which in turn imposes a small hyperfine state splitting. As demonstrated in Fig. \ref{fig:bfield-split} this affects the phase $\phi$ of the atomic superposition state $\Phi$:
\begin{equation}
    \ket{\Phi_\phi}=\frac{1}{\sqrt{2}}(\ket{F=1, m_F=1}+e^{i\phi}\ket{F=1, m_{F}=-1}).
    \label{eq:phase}
\end{equation} 
Stronger hyperfine splitting eventually rotates the phase by an entire revolution and this does not only rotate the polarisation of the emitted photon, but also adversely affects the achievable photon production efficiencies as shown in Fig. \ref{fig:bfield-split}.  A splitting of $1$ MHz corresponds to a field of $1.43$ G, therefore, in the mG regime, as demonstrated in \cite{rempe_cluster}, the phase is close to the desired phase of $\pi$ and the efficiencies are very close to the limit achieved in a perfectly degenerate state configuration. Moreover, strong hyperfine splitting would require a modification of the polarisation of the STIRAP pump beam, such that its polarisation beam matches the polarisation of the emitted photon (cf. Eq. \ref{eq:phase}) where $\phi$ corresponds to the phase of the $\sigma^{\pm}$ beams.
  
\subsubsection{V-STIRAP Photon Production}
\label{sec:app_rb_vst}

\begin{figure}[ht]
    \centering
    \includegraphics[width=\columnwidth]{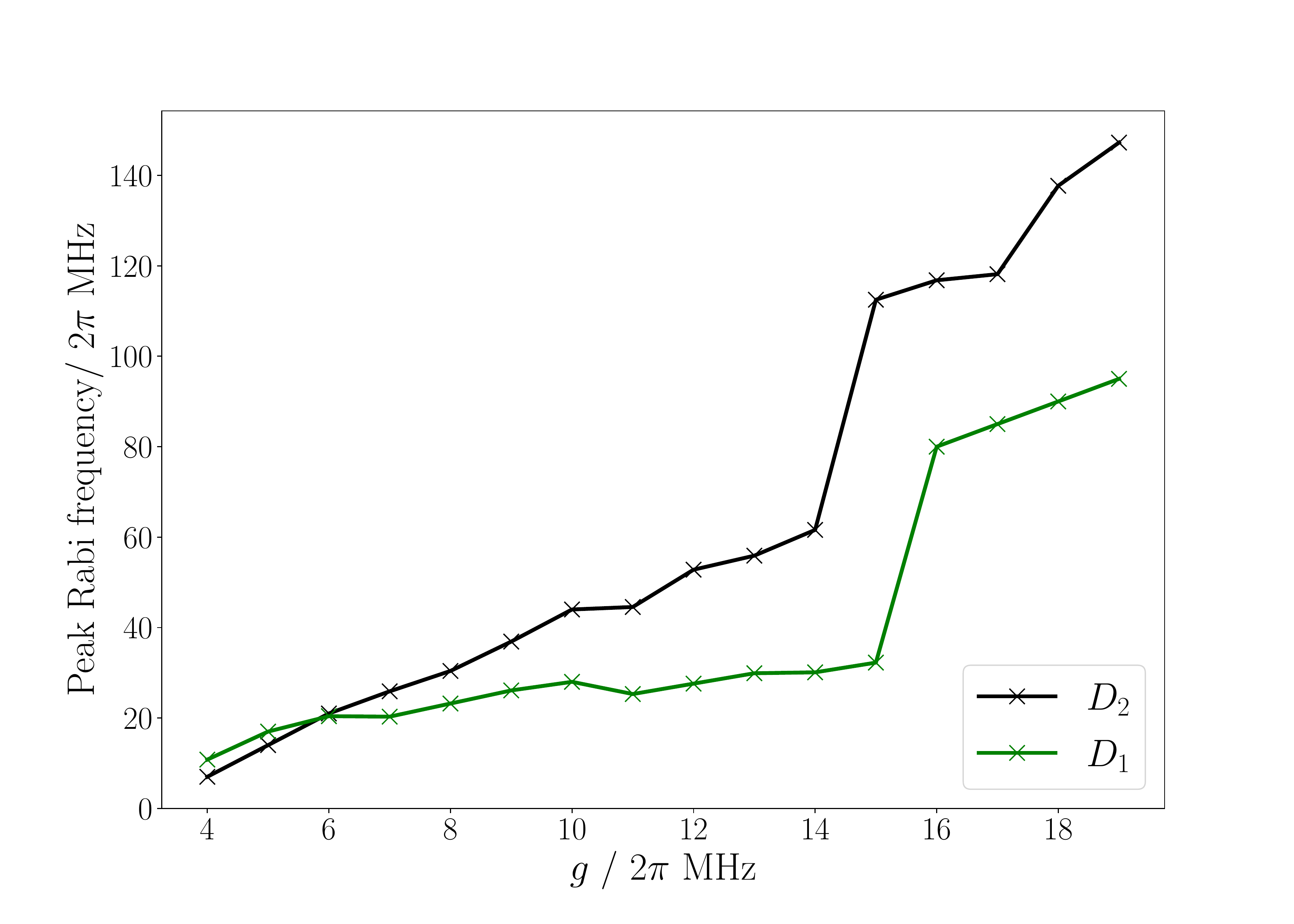}
    \caption{Optimal peak Rabi frequency for the VSTIRAP photon production procedure for various cavity coupling strengths for both the $D_1$ and $D_2$ transition lines.}
    \label{fig:vst-omega-d1-d2}
\end{figure}

As for the idealised system, there exists an optimisation problem at each particular cavity parameter configuration $(g, \kappa)$, to find the optimal peak Rabi frequency and detuning for maximising polarised photon production probability. The optimal peak Rabi frequencies at various atom cavity coupling strengths $g$ and fixed $\kappa=2 \times 2 \pi$ MHz are given in Fig. \ref{fig:vst-omega-d1-d2}. One generally observes an increase with $g$ and $\Omega$ varies approximately linearly for low coupling strengths. Thereafter, $\Omega$ varies more erratically for higher coupling strengths as competing off-resonance effects adversely affect the photon production probability. Detuning the V-STIRAP process further from the atomic resonance had little effect on the efficiency, although detuning the cavity is important insofar as STIRAP re-preparation is performed on the same transitions used for photon production.

\subsubsection{STIRAP Repumping}
\label{sec:app_rb_rep}

The STIRAP re-preparation on either $D_1$ or $D_2$ transitions comprises of a Stokes pulse which is $\pi$ polarised and resonant with the $F=2$ to $F'=1$ transition, as well as a pump pulse which comprises of two $\sigma_{\pm}$ polarised pulses which are resonant with the $F=1$ to $F'=1$ transitions. 

\begin{figure}[ht]
    \centering
    \includegraphics[width=\columnwidth]{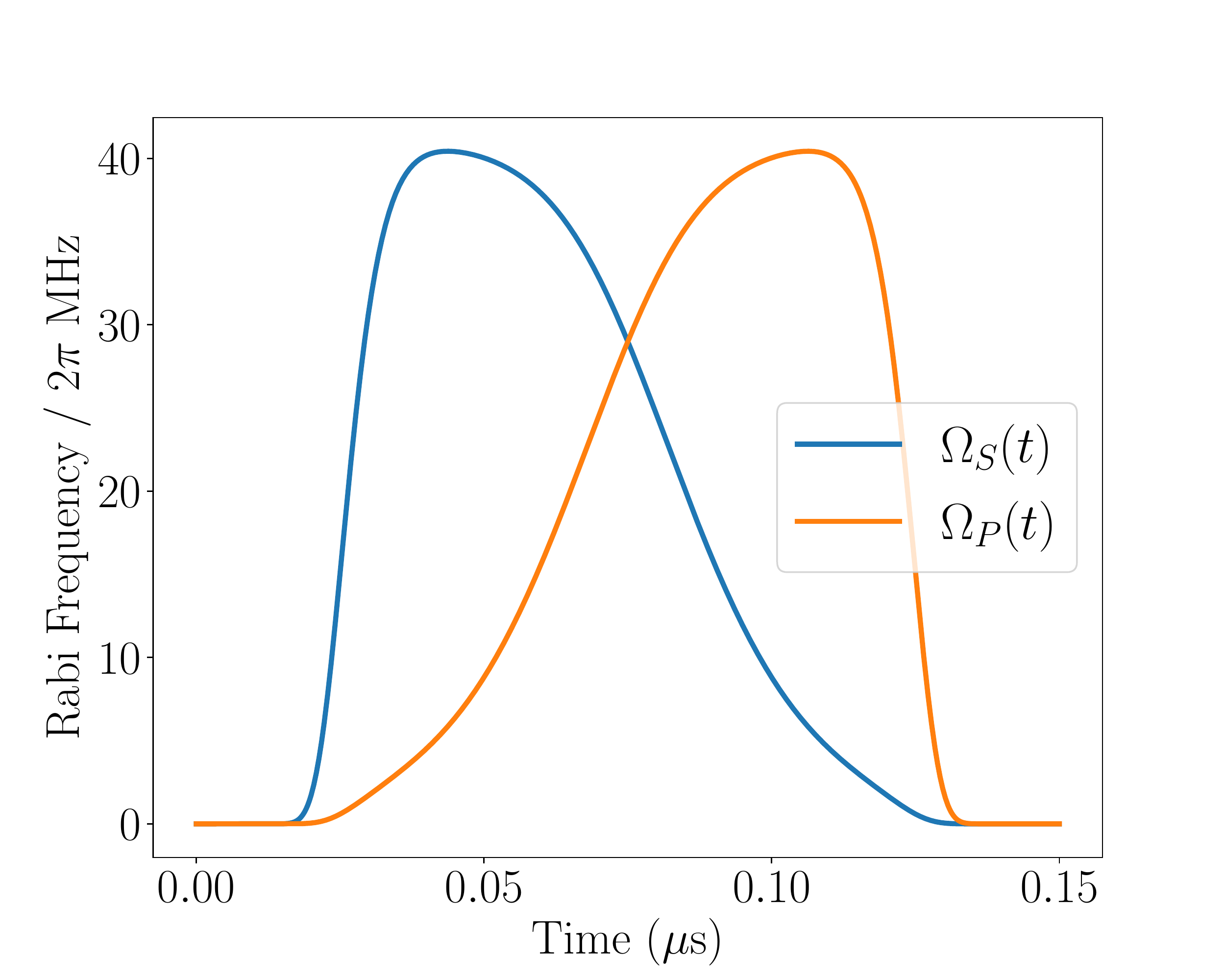}
    \caption{ STIRAP repumping pulses with optimised parameters shown for D1 repumping scheme. Note that although the absolute laser pulse amplitudes differ, the effective Rabi frequencies of the Pump and Stokes pulses for the respective transitions $\ket{\Phi_{\pi}} \xrightarrow{} \ket{F'=1, m_{F'}=0}$ and $\ket{F=2, m_F=0} \xrightarrow{} \ket{F'=1, m_{F'}=0}$ are equal at $41 \times 2\pi$ MHz.}
    \label{fig:rep_real_d1}
\end{figure}

In a real atomic system, the peak Rabi frequency has to be treated as an additional control parameter due to the full atomic structure. We follow the same principles introduced in Sec. \ref{sec:app_ideal_rep} and consider two pulses of constant rms Rabi frequency with a hyper-Gaussian mask with numerically optimised Gaussian shape and peak Rabi frequencies (cf. Eq. \ref{eq:masked_pulses}). Thus, $(n,a, \Omega_{\mathrm{max}})$ are free parameters and $c=0.05$ is chosen such that both pulses exhibit (within the numerical limit of the solver) zero amplitude at $t=0$ and $t=T$. We find the optimal parameters for maximal population transfer efficiency to be as follows: $(n,a)=(6,11)$ and $(\Omega^{D_1}_{\mathrm{max}}, \Omega^{D_2}_{\mathrm{max}}) \approx (41,49) \times 2\pi$. At a fixed repumping time of $150$ ns, the maximal population transfer efficiencies with these pulse shapes are $\approx (0.95,0.83)$. The repumping pulse shapes for $D_1$ are depicted in Fig. \ref{fig:rep_real_d1}.
 It was found that for a time independent pulse detuning, resonant pulses gave the highest population transfer efficiency, but if one seeks to further improve the efficiency, a possible extension is to chirp the pulses by treating the pulse amplitude as a complex parameter. 

\subsubsection{Optical Pumping}
\label{sec:app_rb_op}
\begin{figure}[ht]
    \centering
    \includegraphics[width=\columnwidth]{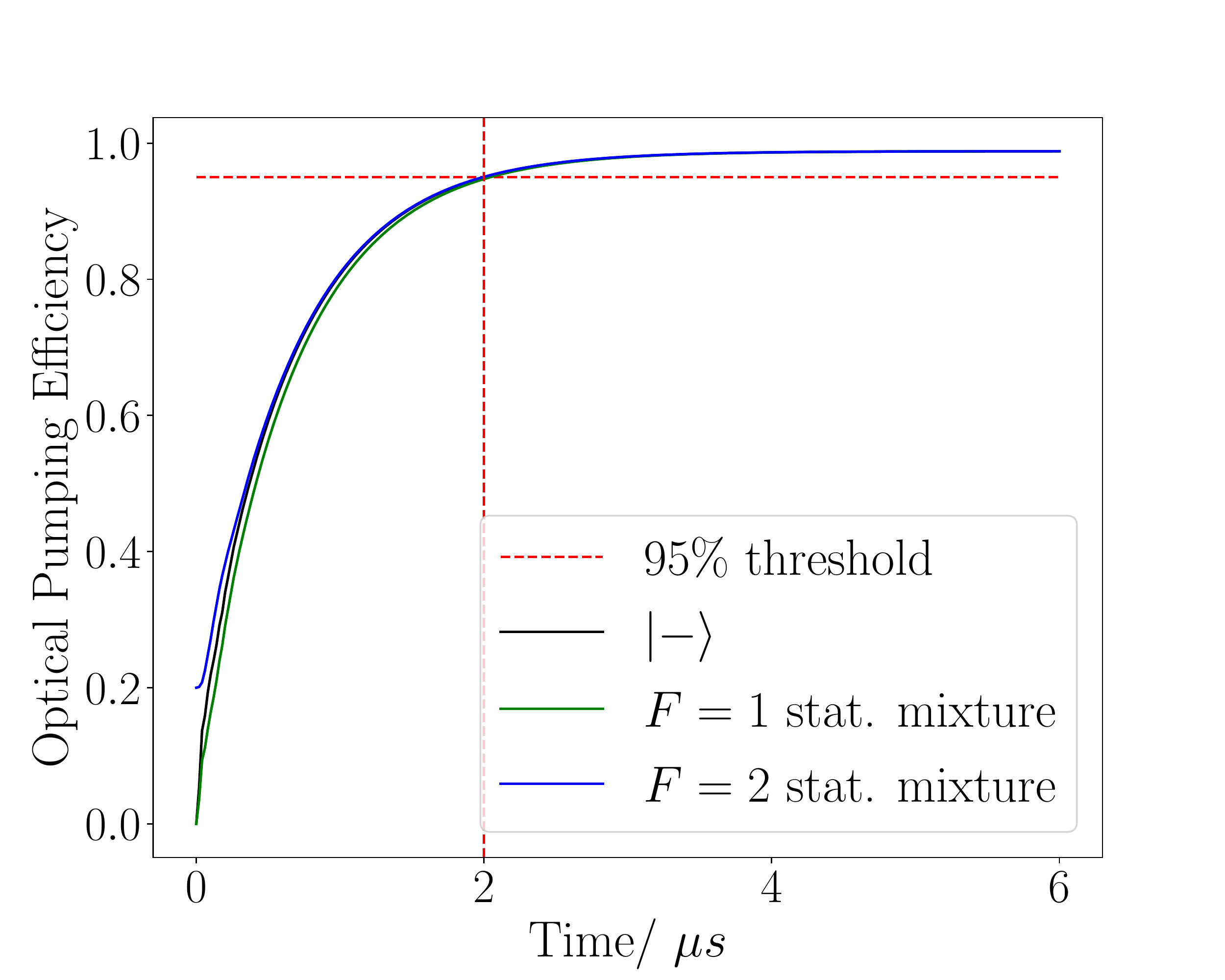}
    \caption{Optical Pumping Efficiency plotted as a function of time for various initial conditions for the $D_2$ transitions. The black trace corresponds to an initial state corresponding to the coherent superposition of the $F=1$ groundstates $\ket{\Phi_{\pi}}$(cf. Eq. \ref{eq:rb_superposition}), the green and blue traces correspond to statistical mixtures of the states of the $F=1$ and $F=2$ levels respectively. Re-preparation efficiency depends on the cost one is willing to incur with respect to the repetition rate, the $95 \%$ threshold is indicated with the dotted red line, and it is approached for times just below $6 \ \mathrm{\mu s}$.}
    \label{fig:op_init_d2}
\end{figure}
Optical Pumping proceeds with two $\pi$ polarised top-hat pulses which have different detunings from the $F=1 \rightarrow F'=2$ and $F=2 \rightarrow F'=2$ transitions, as well as different peak Rabi frequencies. Numerical optimisation with these control parameters was performed to determine the optimal population efficiency to the desired initial state $F=2, m_F = 0$. The simulation results exhibited low pumping efficiencies for resonant pulses due to the creation of two-photon resonances which would repopulate other states of the $F = 2$ level, due to the formation of unwanted dark states formed by individual sub states of $F = 1$ and $F = 2$. Optimal detunings are determined to be $\Delta_1 = 4 $ MHz and $\Delta_2 = -7.5$ MHz, for the $F=1$ and $F=2$ transitions respectively.

Both the $D_1$ and $D_2$ lines were considered for repumping. The optimal peak Rabi frequencies for the $D_1$ transitions were $\Omega_1 / 2\pi = d_{(1, m_i) \rightarrow (2', m_i)}\times 34$ MHz which addresses $F=1$ and $\Omega_2 / (2\pi)= d_{(2, m_i) \rightarrow (2', m_i)}\times 24$ MHz which addresses $F=2$.
The optimal peak Rabi frequencies for the $D_2$ transitions are $\Omega_1 / 2\pi = d_{(1, m_i) \rightarrow (2', m_i)}\times 57.5$ MHz which addresses $F=1$ and $\Omega_2 /2\pi= d_{(2, m_i) \rightarrow (2', m_i)}\times 25.5$ MHz which addresses $F=2$ and where $d_{i,j}$ describe the Clebsch-Gordan coefficients of the transition. The difference in the numerical factors can be attributed to the difference in Clebsch-Gordan coefficients $d_{i,j}$ of the two $F'$ levels.

\subsubsection{Photon Production Cycle}

\begin{figure}[h]
    \centering
    \includegraphics[width=\columnwidth]{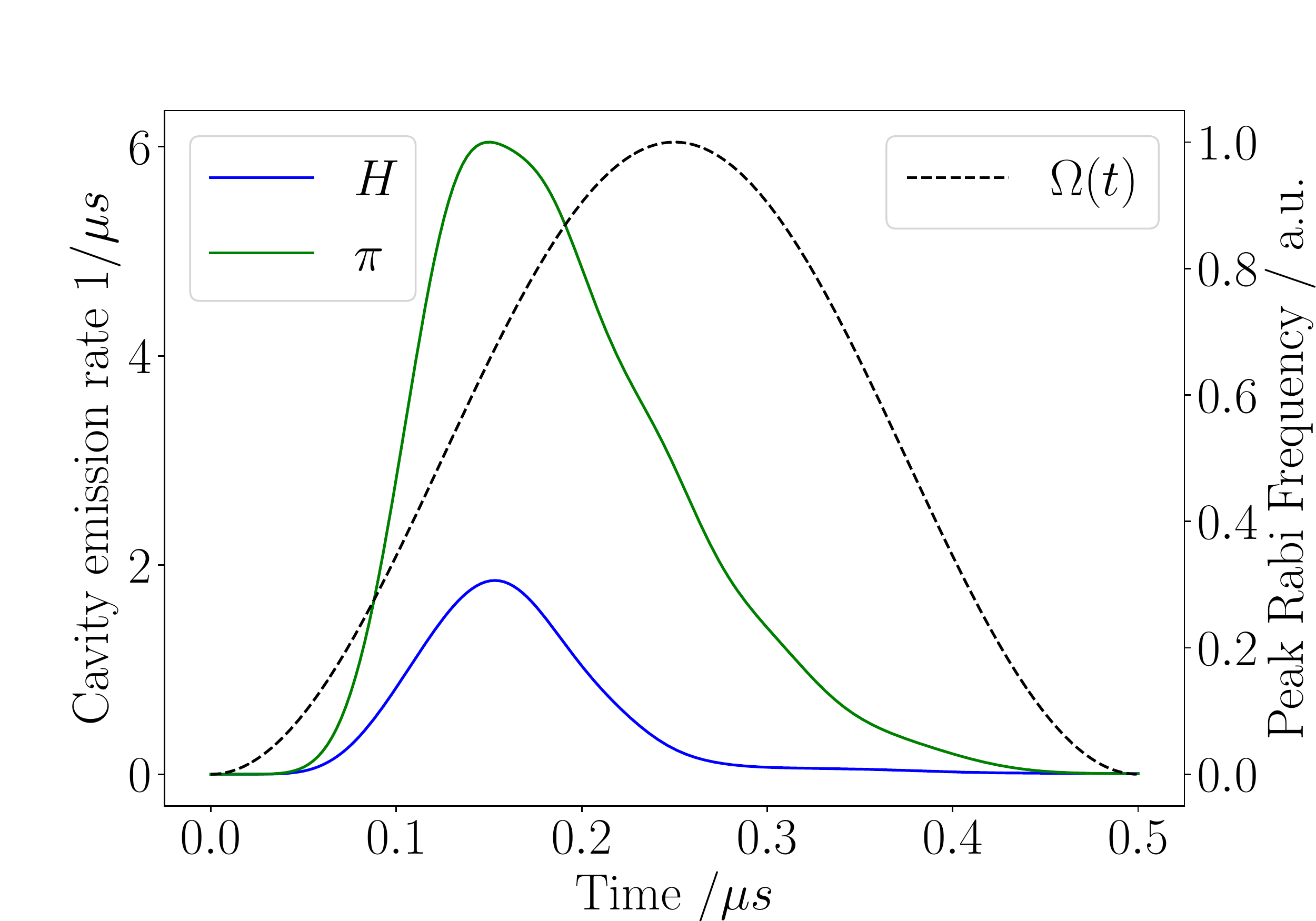}
    \caption{V-STIRAP photon polarisation when the atom is initially falsely prepared into the state $F=2, m_F = -1$, since the $\pi$ transition is no longer dipole forbidden.}
    \label{fig:wrong-state-pol}
\end{figure}

It was important to analyse the distribution of of population after several V-STIRAP \& STIRAP repumping cycles to consider the resulting photon polarisation when a V-STIRAP pulse adresses the atom, after it has been re-prepared to the wrong initial state. In Tab. \ref{tab:pop_cycle} we show the distribution of atomic population in the $F=2$ level after several V-STIRAP and STIRAP cycles. Significant accrual in undesired states adversely affects the efficiency for larger photon chains and since the $\pi$ transitions are no longer dipole forbidden for the $m_F=\pm 1$ states, polarisation errors are introduced, as exemplified in Fig. \ref{fig:wrong-state-pol}. This can however be experimentally mitigated by filtering conditional on photon polarisation at the cavity output.
The optimal parameters for the simulations showcasing expected generation efficiencies for various photon chains, as well as the expected count rates in Fig. \ref{fig:eff_sequence} for an existing cavity design \cite{micro-cav_doherty}, are as follows.  $\Omega_{\mathrm{vst}}\approx105$ and $69\times2\pi$ for V-STIRAP pulses of duration $0.36\text{ \ensuremath{\mu}s}$ and $0.86\text{ \ensuremath{\mu}s}$ respectively, with no detuning with respect to the $\ket{F=2, m_F=0} \rightarrow \ket{F'=1, m_{F'}=0}$ transition and the cavity is resonant with the $\ket{\Phi_{\pi}} \rightarrow \ket{F'=1, m_{F'}=0}$. The optimal pulses are as shown in Fig. \ref{fig:rep_real_d1} for re-preparation via the $D_1$ transition.

\begin{table}[h]
\label{tab:pop_cycle}
\begin{tabular}{l|l|l|l|l|l}
\hline
$F=2, m_F = $ & -2    & -1    & 0     & 1     & 2     \\ \hline
$n = 1$       & 0     & 0.01  & 0.878 & 0.01  & 0     \\ \hline
$n=10$        & 0.121 & 0.078 & 0.44  & 0.078 & 0.121 \\ \hline
$n=100$       &  0.129     &    0.109   & 0.358      & 0.109       & 0.129       \\ \hline
\end{tabular}
\caption{Population after several repumping cycles.}
\label{sec:app_full-scheme}
\end{table}

\end{document}